\documentclass[pre,aps,twocolumn,amsmath,showpacs,amssymb,letterpaper]{revtex4}

\usepackage{hyperref}

\usepackage{graphic x}
\usepackage{here}

\def\cal#1{\mathcal{#1}}
\def\beq{\begin{equation}}
\def\eeq{\end{equation}}
\def\bea{\begin{eqnarray}}
\def\eea{\end{eqnarray}}

\def\eone{\epsilon_{\rm BM}}
\def\etwo{\epsilon_{\rm BS}}
\def\type{\tau}

\def\dc{^{\circ}{\rm C}}
\def\e{\epsilon}
\def\kb{k_{\rm B}}
\def\kt{k_{\rm B} \textnormal{T}_{310}}

\begin{document}
\title{ There and (slowly) back again: \\ Entropy-driven hysteresis in a model of DNA overstretching}
\author{Stephen Whitelam} 
\author{Sander Pronk} 
\author{Phillip L. Geissler}
\affiliation{Department of Chemistry, University of California at Berkeley, and Physical Biosciences and Materials Sciences Divisions, Lawrence Berkeley National Laboratory, Berkeley, CA 94720}
\begin{abstract}
When pulled along its axis, double-stranded DNA elongates abruptly at a force of about 65 pN. Two physical pictures have been developed to describe this overstretched state. The first proposes that strong forces induce a phase transition to a molten state consisting of unhybridized single strands. The second picture instead introduces an elongated hybridized phase, called S-DNA, structurally and thermodynamically distinct from standard B-DNA. Little thermodynamic evidence exists to discriminate directly between these competing pictures. Here we show that within a microscopic model of DNA we can distinguish between the dynamics associated with each. In experiment, considerable hysteresis in a cycle of stretching and shortening develops as temperature is increased. Since there are few possible causes of hysteresis in a system whose extent is appreciable in only one dimension, such behavior offers a discriminating test of the two pictures of overstretching. Most experiments are performed upon nicked DNA, permitting the detachment (`unpeeling') of strands. We show that the long-wavelength progression of the unpeeled front generates hysteresis, the character of which agrees with experiment only if we assume the existence of S-DNA. We also show that internal melting (distinct from unpeeling) can generate hysteresis, the degree of which is strongly dependent upon the nonextensive loop entropy of single-stranded DNA. 
\end{abstract}
\maketitle

\section{Introduction}
DNA in vivo experiences protein-mediated tensile forces large enough to alter the structure and stability of the hybridized state. Single-molecule manipulation techniques permit the study of these deformations through the controlled application of forces and torques~\cite{ten_years}. Intriguingly, DNA pulled along its axis elongates suddenly, by a factor of about 1.7, at a force of 65 pN. Experimental thermodynamics  does not conclusively establish the nature of this overstretched state. The abruptness of elongation suggests that the extended form represents a thermodynamic phase different from B-DNA. The picture of overstretching as a force-induced `melting' to unhybrizided single strands~\cite{bloomfield1,bloomfield2,force_melting, williams, melting_prl} is bolstered by the dependence of the overstretching force on parameters like pH and ionic strength~\cite{williams,bustamante,cocco}; these parameters are known to change the melting temperature of the molecule at zero tension. In addition, the binding of ethidium to DNA increases the overstretching force in a manner consistent with data from thermal melting studies~\cite{williams}. The competing picture, which considers overstretching to be a transition to S-form DNA, a putative hybridized state 1.7 times longer than B-DNA~\cite{bustamante, cocco,s_form1, s_form2, s_form3,zhou}, is motivated by the mechanical properties of overstretched DNA. The latter has a stretch modulus of about $1600$ pN/$0.34$ nm~\cite{cocco}, which exceeds the stretch moduli of both B-DNA ($\sim 1300$ pN/$0.34$ nm) and unhybrizided forms of DNA at comparable forces. However, neither picture can be conclusively ruled out on the basis of these data.
 
Here we show that these thermodynamic data can be used to develop a model of DNA able to discriminate between the {\em dynamics} associated with the two pictures. We have constructed such a model, with a resolution of one basepair, whose kinetics we explore using a Monte Carlo scheme designed to model optical trap pulling experiments. Using as inputs the free energetic properties of DNA in relevant conformations, our model offers thermodynamic and kinetic predictions on experimentally relevant length and time scales. We operate our model in two `modes', corresponding respectively to our interpretation of the force-melting and B-to-S pictures, under conditions designed to model experiments reported in Refs.~\cite{hanbin,rief,cs}. In `force-melting mode' we allow each basepair to instantaneously adopt one of three discrete states: a standard helical state (B), a molten state (M) of two load-bearing but unhybridized strands, and an `unpeeled' state (U) consisting of unhybridized strands of which only one is load-bearing (for illustrations, see Figure~\ref{figmodel}). The latter state is required to model the separation of one strand from another, which may occur when breaks (`nicks') interrupt the sugar-phosphate backbone~\cite{bustamante,cocco,bloomfield1}. Nicks are introduced in
most overstretching experiments as a means of relaxing torsional constraints. In `B-to-S mode' we permit states B, M and U, and in addition assume the existence of an elongated, hybridized form of DNA, the S-state. We specify that this state is 70\% longer than B-DNA, and in equilibrium with the latter at 65 pN; within this framework, the resulting emergent dynamics of our model is largely insensitive to further microscopic details of this putative state of DNA, which we make no attempt to assert or infer.

A typical experiment consists of elongating at constant rate a system consisting of $\lambda$-phage DNA tethered to an optical trap. After the overstretched form of DNA is observed, the elongation is reversed, allowing the molecule to recover its original contour length. A typical stretching-shortening cycle is reversible at low temperature (or high salt concentrations). As ambient conditions are altered so as to destablise the hybridized state, force-extension data become increasingly hysteretic~\cite{s_form1, hanbin}. One notable characteristic of this hysteresis is its {\em asymmetry}: within some temperature range the overstretching plateau displays little or no pulling rate-dependence~\cite{rief,cs}, while the behavior of the shortening transition can vary strongly with pulling rate. This variation becomes more pronounced as temperature is increased~\cite{hanbin}. Thus, one requirement of a viable theory of overstretching is the prediction of a non-hysteretic transition at low temperature and an `asymmetric' hysteretic transition at higher temperature. 

Within our model we find the following. If we assume that there exists an extended double-stranded form of DNA (S-DNA), and further make a minimal set of physically-motivated assumptions about its thermodynamic and mechanical properties, we find that overstretching at low temperature corresponds to the interconversion of two double-stranded forms of DNA (B and S). Because this process involves only local free energy barriers it is relatively rapid, and no hysteresis is observed. At higher temperature unpeeling can occur, starting from nicks in the phosphate backbone. Because of the interplay of the three phases B, S and U, and their distinct mechanical responses, we observe asymmetric hysteresis when unpeeling is considerable. We therefore observe an asymmetric hysteresis that becomes more prominent as temperature is increased. Under conditions designed to model experiments reported in Refs.~\cite{rief, cs}, we observe, following the B-to-S transition, a nonequilibrium high-force unpeeling. The force at which this unpeeling occurs is strongly rate-dependent, and agrees with values obtained in experiment.

In force-melting mode our model reproduces several, principally thermodynamic trends observed in experiment, such as the lowering of the overstretching plateau with increasing temperature~\cite{bloomfield1,bloomfield2,force_melting}. Detailed kinetic predictions, however, do not agree with experimental findings. Unpeeling in our model within a force-melting framework induces a `symmetric' hysteresis, with both shortening and stretching transitions falling out of equilibrium. The height of the overstretching plateau is consequently pulling rate-dependent at all temperatures. The symmetry of the hysteresis derives from the similarity in mechanism for strand detachment and re-annealing: bases open sequentially when strands detach, and close sequentially when strands re-anneal. These processes do not differ sufficiently to render the resulting hysteresis asymmetric. In the presence of the S-phase, by contrast, an asymmetry appears in force-extension data courtesy of the interplay of three mechanically distinct phases (B,U,S). However, we do not discount the possibility that a more thorough treatment of force-melting theory could lead, within our model, to better agreement with experiment. We summarise our findings in Section~\ref{conc}.

We draw considerable inspiration from the insightful theoretical treatment of overstretching of Cocco et al.~\cite{cocco}, although, as discussed later, key features of our model and conclusions differ from those presented in that work.

Although most overstretching experiments are performed upon nicked DNA, and therefore permit unpeeling, our model allows us to make predictions for the case in which unpeeling is suppressed but the DNA remains torsionally unconstrained (for example, un-nicked DNA either in vivo or in vitro with an appropriate tethering arrangement). In this case, molten DNA is the only possible unhybridized form. We find that the nonextensive scaling of the entropy of loops of ssDNA drives an asymmetric hysteresis whose qualitative character depends strongly on the loop exponent $c$ of ssDNA (see Equation~(\ref{loop_factor})). For large values of $c$ ($\sim 2$), the size-dependent free energy barrier associated with a molten loop slows re-annealing sufficiently to generate considerable hysteresis during shortening at typical experimental pulling rates. This hysteresis is a very sensitive function of $c$: for values of $c \sim 1$, we observe little hysteresis associated with melting at typical experimental pulling rates. The predictions of our model for the kinetics of force-induced melting (when unpeeling is suppressed) therefore depend on how $c$ varies with force and temperature. This dependence is currently not known. We speculate that if $c$ increases with temperature (as suggested in Ref.~\cite{c_temp}), we might again observe increasing hysteresis with increasing temperature.
 
This paper is organized as follows. In the following section we discuss the possible origins of the hysteresis that arises during overstretching experiments. In Sections~\ref{thermo0} and~\ref{secdynam} we introduce a simple model for overstretching. We present results obtained from our model in force-melting mode in Sections~\ref{secfm} and~\ref{secfm2}, and discuss its operation in B-to-S mode in Sections~\ref{sec_s2} and \ref{sec_s4}. In Section~\ref{conc} we conclude by suggesting how a more thorough accounting of force-melting theory might lead within our model to better agreement with experiment, and by proposing an experiment that may permit the determination of the loop entropy parameter for DNA under tension.

\section{Entropy-driven hysteresis}
\label{sectwo}

The most striking kinetic effect observed in optical trap pulling experiments is hysteresis at high temperature (e.g. above $30\dc$ at 500 mM NaCl~\cite{hanbin}) in force-extension data for the stretching and subsequent shortening of a pair of initially hybridized DNA molecules. For a large range of temperature this hysteresis is asymmetric: the stretching transition appears to be similar at different temperatures~\cite{hanbin} and pulling rates~\cite{rief,cs}, whereas the shortening transition can exhibit pronounced temperature- and pulling rate-dependence.

Hysteresis signals physical relaxation times in excess of the timescale of an experiment. Common origins of such slow dynamics include large energy barriers due to strong local interactions, large free energy barriers associated with collective reorganization, and the intrinsically slow evolution of long-wavelength motion. The local interactions disrupted by DNA overstretching are noncovalent in character and far too weak to explain the discrepancy between basic rates of basepair opening ($\sim 10^{-8}-10^{-4}$ s~\cite{cocco,bubbles0,bubbles1}) and apparent relaxation rates (seconds to minutes). The implicated importance of collective dynamics, however, is highly unusual for systems macroscopic in only one dimension, because in one dimension the interfaces between domains do not grow with increasing domain size. Long-ranged correlations are difficult to accumulate if the penalty for destroying them is microscopic, as is the case, for example, when melting a single turn of the double helix. Possible explanations for the hysteresis observed in DNA overstretching are thus few in number and special in character.

In this paper we consider two such explanations. The dominant mechanism through which hysteresis arises in experiments performed to date is most likely the `unpeeling' of one strand from the other, beginning from nicks in the phosphate backbone. This has been suggested by other authors~\cite{bustamante,cocco,bloomfield1}; here, we investigate this mechanism in a framework that incorporates thermally-driven dynamics occurring on a broad range of timescales. Within our model we indeed observe hysteresis when we allow unpeeling in the presence of nicks. This hysteresis arises from the slow, long-wavelength progression of the unpeeled `front' as it advances (or retreats),  basepair by basepair, over a macroscopic distance. In the presence of the S-phase this hysteresis is asymmetric. For moderate temperatures unpeeling begins at high forces, at or above the force of the B-to-S transition. The slow conversion of S- to U-DNA at high forces is not readily seen in force-extension plots, because the mechanical responses of these phases in the high-force regime are so similar (see Section~\ref{thermo0}). By contrast, the slow conversion of U-DNA to B-DNA upon relaxation shows a strong force-extension signature, because the mechanical responses of these states in the lower-force regime are very different. We present results obtained using the sequence of $\lambda$-phage DNA, but sequence heterogeneity is not required to generate such hysteresis within our model. In the absence of the S-state (with or without internal melting), unpeeling-generated hysteresis is symmetric, because the mechanism of the advance and retreat of the unpeeled front is the same for stretching as for shortening.

Because molten DNA is thermodynamically unstable with respect to unpeeled DNA~\cite{cocco}, in the presence of nicks unpeeled DNA predominates over molten DNA under all conditions. We present simulation results corresponding to this case, relevant to most experiments, in Section~\ref{sec_nicks}. However, our model allows us to make predictions for overstretching in the absence of nicks, where the only possible form of unhybridized DNA is the molten state. This scenario is of considerable interest in biological contexts. We find that when melting is substantial, there exists a possible alternative mechanism for generating hysteresis, originating in the conformational entropy of `bubbles' of molten DNA. This conformational entropy induces long-ranged correlations that may persist for times comparable to the shortening protocol. For the remainder of this section we shall outline the mechanism by which melting can generate hysteresis: we explore the effect of this mechanism upon overstretching in Section~\ref{sec_nonicks}.

The entropy associated with molten regions of tension-free DNA does not grow linearly with the number of consecutive unhybridized bases. A `bubble' of $n$ unpaired bases can be regarded as a loop of single-stranded DNA of length $2n$, whose entropy has a contribution~\cite{poland, fisher} that grows sub-linearly with $n$:
\beq
\label{loop_factor}
\Delta S_{\rm loop}(n) = - \kb \, c \ln (2n).
\eeq
This nonextensive correction reflects the loss of entropy induced by constraining a freely-fluctuating strand to end at its origin (thereby forming a loop). Pinching a bubble in its middle incurs a statistical penalty $\sim n^{-c}$ (see below). A positive loop exponent $c$ therefore favours large bubbles over multiple smaller ones. It has been known for decades that this effect leads to a `sharpening' of the melting transition~\cite{poland}.

The nonextensive entropy of ssDNA also has a direct consequence for the kinetics of melting and rehybridization. Assuming bulk phase coexistence between helical and ssDNA, the probability of nucleating a single basepair of ssDNA within the helix is proportional to
\bea
\label{loop_one}
w_{\rm nucleate}&=& \sigma_0 \rm{e}^{ \Delta S_{\rm loop}(1)/\kb } \nonumber \\
&=& \sigma_0  \, 2^{-c},
\eea
where $\sigma_0 \sim 10^{-5}$ is the local penalty for boundaries between helical and molten DNA~\cite{BZ}. Conversely, the probability of nucleating a single basepair of helical DNA in the centre of a loop of $n$ basepairs is proportional to
\bea
\label{loop_two}
w_{\rm pinch} &=& \sigma_0 \rm{e}^{ 2 \Delta S_{\rm loop}(n)/\kb- \Delta S_{\rm loop}(2n)/\kb} \nonumber \\
&=& \sigma_0  \, n^{-c}.
\eea
From Equation~(\ref{loop_one}) we see that the free energy penalty for the nucleation of ssDNA is purely local, being insensitive to the size of the helix within which it attempts to nucleate. By contrast, from Equation~(\ref{loop_two}) we see that the free energy barrier for the nucleation of double-stranded (dsDNA) within a region of ssDNA, which we call loop `pinching', grows with the size of the loop in question. Pinching large bubbles is very difficult. Since the closing of a bubble from its ends is a slow process (for the same reason that unpeeling is slow), large molten bubbles can resist complete closure for long times even when thermodynamically unstable. The degree of resistance to pinching depends strongly on the value of the loop exponent $c$. Estimates for  $c$ under zero tension range from $\sim 1.8$~\cite{fisher} to $\sim 2.1$~\cite{kafri}.  

\begin{figure}[tb] 
   \centering
         \includegraphics[width=3in]{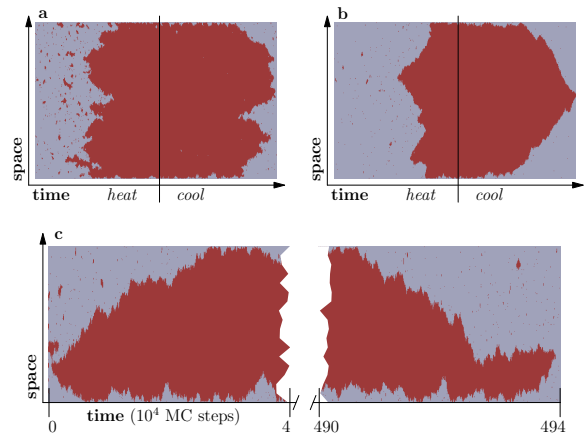} 
   \caption{\label{figheating}
Illustration of the asymmetry of melting-hybridization dynamics. We show microscopic configurations (vertical) as a function of time (horizontal) for a 500-basepair toy model of DNA having two phases, hybridized (blue) and molten (red). The sequence is homogeneous. {\bf Top panel}: heating-cooling cycle for $c=1$ ({\bf a}) and $c=2$ ({\bf b}). We increase the temperature linearly with time until well above coexistence, and then reverse the protocol until the original temperature is recovered. Molten bubbles grow via nucleation, and close via nucleation when the parameter $c$ is small (left panel). When $c$ is large (right panel), cooling is sufficiently rapid that bubbles are constrained to close via domain wall drift. This asymmetry can give rise to melting-driven asymmetric hysteresis in pulling simulations, in which the varying control parameter is tension rather than temperature (see Section~\ref{sec_nonicks}).
{\bf Bottom panel (c):} equilibrium dynamics for $c=2$ at a fixed temperature near coexistence. With no external driving, dynamical symmetry is restored.}
   \end{figure}

We can illustrate the difference in mechanism between melting and reannealing using the simple system shown in Figure~\ref{figheating}. Here we show a Glauber Monte Carlo simulation of a one-dimensional caricature of a 500 basepair fragment of DNA. The model consists of a blue (`hybridized') phase having energy per unit length $\epsilon<0$, and a red (`molten') phase having entropy per unit length $T S_{\rm m}>0$ ($T$ is temperature). In addition, the molten phase possesses a nonextensive entropy $-c T \ln (2 N)$, $N$ being the number of consecutive molten basepairs. For simplicity we ignore sequence heterogeneity. Starting from an initially hybridized state, we subject this system to a heating-cooling cycle. As temperature is increased, ssDNA becomes favoured thermodynamically, and begins to nucleate within the helix. At sufficiently high temperature, when most of the system is unhybridized, we reverse the heating protocol. As temperature falls, hybridized DNA becomes increasingly favoured thermodynamically, and sections of the system attempt to anneal. For small values of $c$ (panel {\bf a}) the free energy barrier to loop pinching is small; molten bubbles can close by pinching. In this regime melting-hybridization dynamics is roughly symmetric. For larger values of $c$ (panel {\bf b}) large loops resist pinching, and close in general slowly from their from their ends. Provided the heating-cooling cycle is sufficiently rapid, melting-hybridization dynamics is {\em asymmetric}. Dynamical symmetry is of course restored at equilibrium (panel {\bf c}), where the critical nucleus for melting spans the entire length of the molecule.

This melting-hybridization asymmetry has consequences for the kinetics of conformational changes of DNA whenever melting is appreciable. In the context of DNA overstretching, we expect melting to be substantial only when unpeeling is suppressed (for example, in the absence of nicks). If  this is so, we find that melting likely plays a role in both pictures we have described for the overstretching transition. In the force-melting picture the elongated form of DNA is unhybridized. Within the B-to-S picture, at hight temperature melting pre-empts the B-to-S transition. If we adopt the zero-tension form of the loop factor, Equation~(\ref{loop_factor}), and take $c \sim 2$, we indeed find in simulations of overstretching that when melting is appreciable, considerable hysteresis is generated during the shortening transition at typical loading rates, regardless of whether we assume the existence of S-DNA. In order to fully understand the kinetic predictions of each picture of overstretching (in the absence of nicks), it is necessary to understand the behavior under tension of the the nonextensive entropy of ssDNA.

There is currently no experimental data describing the force dependence of the nonextensive entropy of ssDNA. Calculations that neglect the excluded volume of single strands (see Appendix A) suggest that in the presence of a force $f$ the nonextensive loop entropy takes the form
\beq
\label{loop_factor2}
\Delta S_{\rm loop}(n,f) = - \kb \, c \ln (2n) + \kb \, \ln g(f).
\eeq
Although this entropy as a whole is tension dependent, the exponent $c$ is not. The correction factor $g(f)$ does not depend on $n$ and can be absorbed into the local surface tension. The penalty for loop-pinching, $w_{\rm pinch}(f) = \sigma_0  \, g(f) \, n^{-c}$, {\em relative} to loop nucleation, $w_{\rm nucleate}(f) = \sigma_0  \, g(f) \, 2^{-c}$, is essentially unchanged by force. Note that at forces typical of overstretching, 65 pN, $g(65 \, {\rm  pN}) \sim 10^3$, resulting in a renormalized boundary parameter of $\sigma_r(65\, {\rm  pN}) \equiv \sigma_0 \, g(65 \,{\rm  pN}) \sim 10^{-2}$ and a reduced `cooperativity' of melting. 

Calculation~\cite{rouzina1} suggests that when the excluded volume of ssDNA is accounted for, the correction $g(f)$ becomes $n$-dependent. Intuition further suggests that $c$ may be reduced by tension, to the extent that loops interact less strongly with a taut backbone than with a floppy one~\cite{kafri2}. In simulations we vary $c$ to identify the regime at which hysteresis sets in. We find that a value of $c \sim 1$ would yield a largely non-hysteretic force-induced melting transition at experimental loading rates.

\section{Model thermodynamics}
\label{thermo0}
\subsection{Modeling force-induced melting}
\label{thermo}
\begin{figure*}[ht] 
 \begin{center}
 $\begin{array}{c}
\includegraphics[width=13.5cm]{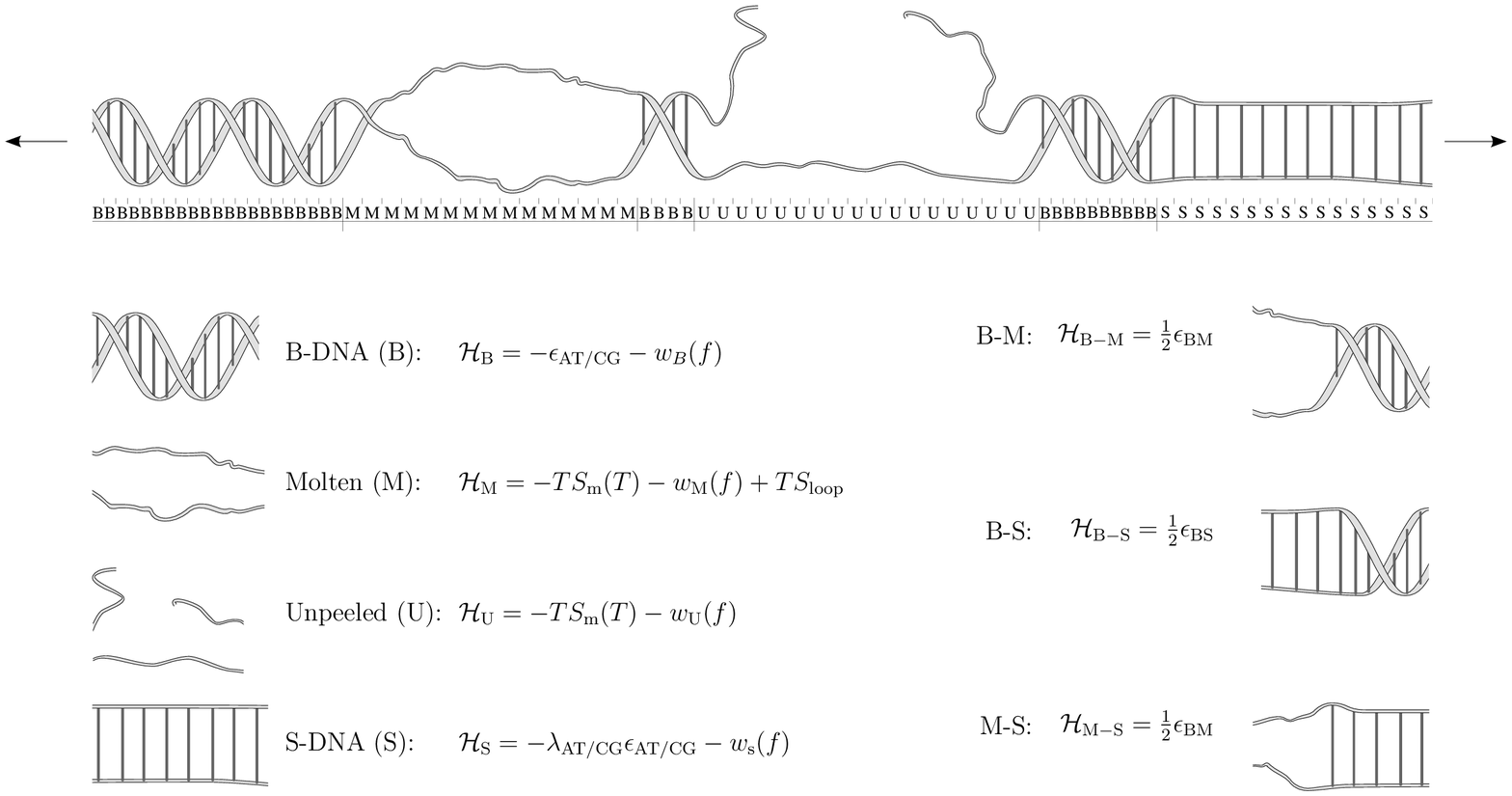} 
\end{array}$
\end{center}
   \caption{\label{figmodel}
Illustration of our model. The model is one dimensional, and consists of $N$ `sites' corresponding to coarse-grained basepairs. Each site is of type CG or AT according to the sequence of $\lambda$-DNA (sequence-dependence not shown) and at any instant will correspond to one of states B, M, U or S.  We show an example configuration (top) of a fragment of the model, illustrating the conformation of DNA that we imagine corresponds to a given configuration of sites (sequence of letters). Shown below the configuration are schematic pictures of each state, together with their extensive and nonextensive free energies [see Equations~(\ref{hamiltonian}),~(\ref{nonex}) and~(\ref{hamil_s})]. On the right we show free energies associated with interfaces between different states. We operate our model either in force-melting mode (in which we permit states B, M and U) or B-to-S mode, in which case we permit in addition the hypothetical S-state of DNA. }
   \end{figure*}

 To construct a model of overstretching we imagine coarse-graining a DNA molecule into sites spanning the distance along the backbone between adjacent basepairs, `blurring out' details on smaller scales. We illustrate this idea in Figure~\ref{figmodel}. Each site (`basepair') is fixed to be of type AT or CG (we assume fully complementary strands), and at any instant adopts one of the states B, M, or U. In the following subsection we shall introduce a fourth state, S. Such discrete models of biopolymers have a long history, beginning with the two-state model of Bragg and Zimm~\cite{BZ}, used in the 1950s to study the melting transition of homopolymers. More elaborate discrete models of conformational transitions in DNA have since been considered, including the model of SantaLucia~\cite{santa}, and a static three-state model of overstretching~\cite{cocco}. These models allow one to make predictions on time and length scales (seconds, thousands of basepairs) greatly in excess of those available to molecular dynamics techniques (microseconds, tens of basepairs)~\cite{entropy2}, albeit at a much lower level of detail.
   
To model the fragments of $\lambda$-phage DNA used in experiments we study a system of size $N=4.1 \times 10^4$~\cite{hanbin} or $N=4.5 \times 10^3$~\cite{rief,cs} as appropriate, with sites fixed to be of type AT or CG according to the sequence of $\lambda$-phage DNA~\cite{lambda}. We denote site location by the variable $i \in [1,N]$, site {\em type} by the variable $\type_i \in \{$AT,CG$\}$, and the {\em state} of a given site by the variable $\alpha \in \{ $B,M,U,S$\}$. We shall use the terms `site' and `basepair' interchangeably.

The equilibrium probability that a basepair adopts a given state $\alpha$ in our model is set by the free energetic properties of DNA in the relevant conformation. We shall quote energies in units of $\kb T$ at 310K, which we write as $\kt$. Recall that $\kt \approx$ 0.617 kcal/mol $\approx 4.265$ pN nm. The dynamics of our model, discussed in the following section, follows naturally from these statistical weights via a standard Glauber Monte Carlo algorithm.

In the remainder of this subsection we shall focus on the force-melting theory of DNA overstretching, and allow only states B, M and U. In Section~\ref{sec_s} we shall model the putative S-state of DNA.

We call the logarithm of the statistical weight in our model, $\cal{H} = -\kb T \ln W$, the `Hamiltonian': it is a free energy in which we imagine atomistic degrees of freedom have been integrated out. We separate the Hamiltonian into local and nonlocal pieces, $\cal{H}=\sum_{i=1}^N \left( \cal{H}_i^{\rm l} +\cal{H}_i^{\rm nl}\right)$. The local piece contains the extensive free energy of each basepair, as well as interactions between adjacent basepairs. For a given site $i$, 
\bea
\label{hamiltonian}
\cal{H}_i^{\rm l} =&-&\left[\e_{\type_i} +w_{\rm B}(f) \right]  \delta_i^{\rm B} \nonumber \\ 
&-&\left[T S_{\rm m}(T) + w_{\rm M} (f)\right] \, \delta_i^{\rm M}  \nonumber \\ 
&-&\left[T S_{\rm m}(T) + w_{\rm U} (f)\right] \, \delta_i^{\rm U}  \nonumber \\ 
&+&\frac{1}{2} \eone \,  \sum_{\gamma={\rm M}, {\rm U}} \left( \delta_i^{\rm B} \delta_{i+1}^{\rm \gamma}  +\delta_i^{\rm \gamma} \delta_{i+1}^{\rm B} \right).
\eea
We have introduced the binary variables $\delta_i^{\alpha}$, which take the value 1 when site $i$ assumes the state $\alpha$, and 0 otherwise. The variable $f$ is the molecular tension. 

The nonlocal piece accounts for the nonextensive entropy of molten loops, and for a loop beginning at site $i$ reads
 \beq
 \label{nonex}
\cal{H}_i^{\rm nl} = -T \sum_{n=1}^{N-i} \Delta S_{\rm loop}(n,f) \, \tilde{\delta}_{i-1}^{\rm M} \tilde{\delta}_{i+n}^{\rm M} \prod_{s=0}^{n-1} \delta_{i+s}^{\rm M}.
 \eeq
The complementary binary variables $\tilde{\delta}_i^{\alpha}\equiv 1-\delta_i^{\alpha}$ take the value 0 when site $i$ assumes the state $\alpha$, and 1 otherwise. The nonextensive loop entropy penalty can be calculated, for example, within the freely-joined chain model (ignoring volume exclusion), as~\cite{rouzina1}  $-T \Delta S_{\rm loop}(n,f)\equiv \kb T \, c \ln(2n)-\kb T \ln g(f)$. The force-dependent function $g(f)$ provides an approximate means of accounting for the change in loop entropy with force, and reads~\cite{rouzina1}
\beq
\label{gee_0}
\frac{2}{g(f)} =  {\cal L}(f_{\rm r}) \left({\cal L}(f_{\rm r})+\frac{2}{f_{\rm r}^2}-\frac{2}{f_{\rm r}} \coth (f_{\rm r})\right)^{1/2}.
\eeq
Here $f_{\rm r} \equiv f P_{\rm ss }/(\kb T)$ is a reduced force, $P_{\rm ss } \approx 0.7$ nm is the persistence length of ssDNA, and ${\cal L}(x)\equiv 1- [\coth (x) - 1/x]^2 $.

In what follows we shall define the parameters of Equation (\ref{hamiltonian}). The first three lines model the extensive free energies of homogeneous B, M and U phases, respectively. The fourth line describes the unfavourable free energy associated with interfaces between helical and ssDNA~\cite{BZ}. The parameters of the Hamiltonian associated with each phase derive from molecular free energies of two types: force-independent and force-dependent. We shall describe the force-independent parameters first.

{\em Force-independent free energies.} B-DNA possesses both basepairing and stacking interactions, which we denote by $\e_{\type_i}$; the variable $\type_i \in \{{\rm AT},{\rm CG} \}$ specifies the {\em type} of basepair $i$. We estimate these interactions from thermal melting studies. We assume that these energies depend only upon the type $\tau_i$ of basepair $i$ and not on the type of the neighbouring sites $i\pm1$. This frequently-made approximation can be regarded as a `renormalization' of the neighbour-dependent stacking energies of more detailed models~\cite{santa}, by dividing into two sets interactions containing 1) any A or T bases, or 2) any C or G bases. This level of detail is sufficient to capture the essence of sequence heterogeneity, namely the extra stability against de-hybridization of CG over AT basepairs, which in free energy terms is roughly 2 $\kb T_{310}$. It also allows us to capture the variation with AT content of the melting temperature of DNA. We fix the values of the energies $\e_{\rm CG}$ and $\e_{\rm AT}$ using values from the literature~\cite{bloomfield1, cocco, santa} (we assume that the entropy of melting liberated per basepair is $S_{\rm m}=12.5 \, \kb$; see below). We take
\bea
\e_{\rm AT}=13.17+0.19\, \ln(M/0.150) \nonumber \\
\e_{\rm CG}= 15.28+0.19\, \ln(M/0.150),
\eea
in units of $\kt$. Here $M$ is the molar NaCl concentration (we quote the salt-dependent correction from~\cite{cocco}). We choose as a salt concentration either 150 mM NaCl (so as to model the experiments of Refs.~\cite{rief,cs}) or 500 mM NaCl (in order to model experiments reported in Ref.~\cite{hanbin}).

In addition, in B-DNA there exists an interaction between each nucleotide and a nucleotide on the opposing strand four basepairs distant. This interaction is lost upon disruption of the helix~\cite{BZ,pb,pbd}. We account for this effect by assigning an interaction parameter $\e_{\rm BM} \approx 11.51 \,  \kb T $~\cite{BZ,cocco} to neighbouring B- and non-B sites (fourth line of Equation (\ref{hamiltonian})). This interaction corresponds to a {\em junction penalty}, familiar from Bragg-Zimm~\cite{BZ} models, of $\sigma_0 \equiv \exp \left(-\eone/(\kb T) \right) = 10^{-5}$.

The statistical weight of the M-phase receives a contribution from the extensive entropy per basepair (bp) liberated upon melting (under zero tension), $S_{\rm m}(T)$, which we take to be~\cite{bloomfield1,bloomfield2}
\beq
\label{entropy}
S_{\rm m}(T)=S_{\rm m}(T_{\rm m})+\frac{\Delta C_p}{2} \frac{T_{\rm m} }{T} \left(\frac{T_{\rm m}-T}{T_{\rm m}} \right)^2,
\eeq
where $S_{\rm m}(T_{\rm m}) = 12.5 \, \kb$. The second term (in which we have absorbed both entropic and enthalpic contributions~\cite{bloomfield2}) derives from the change in heat capacity upon melting, quoted as $\Delta C_p \approx 60 \pm 10$ cal/(mol$\cdot$K$\cdot$bp)~\cite{bloomfield2}. We set $\Delta C_p = 60$ cal/(mol$\cdot$K$\cdot$bp). The specific heat correction to the free energy increases from zero at the melting point to about $1 \kt$ at 0$\dc$. The experimental melting temperature $T_{\rm m}$ for $\lambda$-phage DNA (which has $\sim$50\% CG content) has been fit by~\cite{bloomfield1,bloomfield2} $T_{\rm m}(\dc) \approx 104.4 +16.89 \log_{10} M$. This formula gives a melting temperature of about 90$\dc$ at 150 mM, in reasonable agreement with the zero-force melting point of our model (about 87$\dc$ for $c=2$; data not shown).
 
We assume the unpeeled state (U) to be unhybrizided DNA consisting of one load-bearing single strand; the other strand is free to fluctuate~\cite{cocco}. We take its extensive entropy to be that of M-DNA, and we assume that U-DNA has no nonextensive entropic correction. In our model, any unhybridized site connected to at least one nicked site by an unbroken sequence of unhybridized sites is defined to be unpeeled; if not so connected, it is defined to be molten. We assume that there is no energetic barrier between the M and U states~\cite{cocco}. We set the energy of U-B junctions equal to those of M-B junctions.

{\em Force-dependent free energies.} The terms $w_{\alpha}(f)$ in the first three lines of Equation (\ref{hamiltonian}) describe the elastic free energies of the B, M and U states of DNA. One can determine these energies by integrating, as a function of pulling force, the experimentally-measured extension per basepair $x_{\alpha}(f)$ of each state~\cite{cocco,bloomfield1}, i.e. 
\beq
w_{\alpha}(f) \equiv \int_0^f df' \, x_{\alpha}(f').
\eeq
For forces of interest, the extension per basepair of B-DNA as a function of force $f$ is given approximately by an extensible worm-like chain model~\cite{oddjob},
\beq
\label{bform}
x_{{\rm B}}(f) = \bar{x}_{\rm B} \left(1-\frac{1}{2} \left( \frac{\kb T}{f P_{\rm B}} \right)^{1/2} +\frac{f}{S_{\rm B}} \right),
\eeq
with contour length $\bar{x}_{\rm B}= 0.34$ nm/bp, persistence length $P_{\rm B}=50$ nm, and stiffness $S_{\rm B} = 1300$ pN. We bound the above expression (which breaks down at very small forces $f < 0.1$ pN) from below by zero.

We take the mechanics of M-DNA to be that of two pieces of load-bearing ssDNA, for which we adopt the parameterization of Ref.~\cite{cocco}, 
\beq
\label{ssDNA}
x_{ \rm ss}(f) = \bar{x}_{\rm ss} \left( \frac{a_1 \log(f/f_1)}{1+a_3 \textnormal{e}^{-f/f_2}}-a_2-\frac{f}{f_3}\right).
\eeq
Here $\bar{x}_{\rm ss} = 0.34$ nm, $a_1=0.21$, $a_2=0.34$, $a_3=2.973 +0.5129 \ln M$, $f_1=3.7 \times 10^{-3}$ pN, $f_2 = 2.9$ pN, and $f_3=8000$ pN. This formula accounts for the increase in persistence length of ssDNA under low salt conditions. We bound this expression from below by zero. We find that the extensible freely-joined chain fit to ssDNA~\cite{smith}, which does not account for the increase in its persistence length at low salt concentrations, leads to numerically different melting forces at nonzero tension, but does not affect the qualitative predictions of our model. We neglect the effects of secondary structure, which for load-bearing strands we expect to be small. 

We assume that M-DNA consists of two noninteracting, load-bearing strands, and so we take $x_{\rm M}(f) = x_{\rm ss}(f/2)$ and $w_{\rm M}(f) = 2 w_{\rm ss}(f/2)$. We asssume that U-DNA consists of a single load-bearing strand~\cite{cocco}, and so take $x_{\rm U}(f) = x_{\rm ss}(f)$ and $w_{\rm U}(f) = w_{\rm ss}(f)$.

In principle there is a force-dependent contribution to the entropy of ssDNA arising from the `pulling out' at high forces of its backbone degrees of freedom. This effect is negligibly small, however, as may be shown within the freely-jointed chain model of ssDNA.
\begin{figure}[tb] 
   \centering
         \includegraphics[width=3.2in]{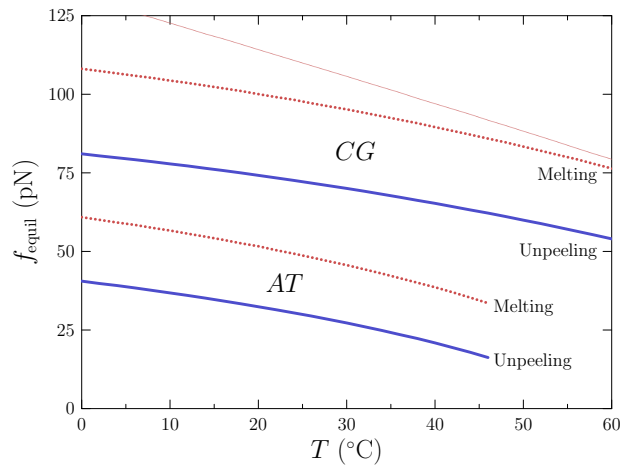} 
   \caption{\label{fignrg} Force (vertical axis) at which M- (red) and U-DNA (blue) are in equilibrium with B-DNA, as a function of temperature (horizontal axis). This calculation is based on the extensive free energies of the Hamiltonian, given in Equation~(\ref{hamiltonian}). The lower pair of bold curves refers to AT sites; the upper pair refers to CG sites. The thin line was computed in the absence of the specific heat correction shown in Equation~(\ref{entropy}). This correction markedly lowers the overstretching force at low temperature, giving much better agreement with experiment. With no free energetic barrier between M and U, molten DNA is always thermodynamically unstable to unpeeling~\cite{cocco}. In the presence of nicks (the usual experimental situation) we therefore find unpeeled (U) DNA to be the predominant form of unhybridized DNA.  }
   \end{figure}

Finally, we approximate the nonextensive loop entropy of ssDNA via Equations~(\ref{nonex}) and~(\ref{gee_0})~\cite{rouzina1}. This approximation neglects the excluded volume of single strands. In view of this, we regard the loop exponent $c$ as an unknown and vary it between 0 and 2.1.

We shall examine the predictions of our model both in the presence and the absence of nicks. In the presence of nicks (the usual experimental situation) we permit unpeeling. Within our model we find that molten DNA is always thermodynamically unstable to unpeeling. This is so by virtue of the unpeeled state being longer at a given force than the molten state (a single load-bearing strand extends more than do two strands each bearing half the load). A similar conclusion was reached in Ref.~\cite{cocco}. The free energy of U-DNA can fall below that of the B-state at a lower force than can the free energy of M-DNA. There is therefore a range of $T$ and $f$ for which unpeeling can cross CG regions (and therefore proliferate over long distances), while melting cannot.

We demonstrate the instability of melting to unpeeling in Figure~\ref{fignrg}, in which we show the force (as a function of temperature) at which B-DNA is unstable to unpeeling and to melting. For nicked molecules the predominant form of unhybridized DNA is therefore unpeeled. We shall examine the predictions of our model in this case in Sections~\ref{sec_s2} (B-to-S mode) and~\ref{secfm} (force-melting mode).

\subsection{Modeling S-DNA}
\label{sec_s}

To model the picture of overstretching as a transition to an elongated form of dsDNA, we extend our model to include a fourth state, S-DNA~\cite{bustamante, cocco,s_form1, s_form2, s_form3,zhou}. S-DNA, whose existence is debated, is naturally less well-characterized than either helical or molten DNA. We must therefore make several assumptions about its thermodynamic and mechanical properties. However, within our model a minimal set of physically-motivated assumptions results in a kinetics whose qualitative features are robust to changes in model details. Our key assumptions in this section and the section following are 1) that S-DNA exists; 2) that it is double-stranded (and so its free energy possesses no long-ranged entropic component); 3) that it is 70\% longer than B-DNA, and 4) that it is in thermal equilibrium with the latter at 65 pN. Within this framework the qualitative kinetic details that emerge depend very weakly on the precise numerical values of our S-DNA parameters. 

The B-to-S picture must be considered tentative until there exists a direct observation of this state of DNA. The force-melting picture, by contrast, is a first-principles approach with considerable predictive power, building on well-known properties of helical and molten DNA. In this study we do not propose or defend a likely atomistic model for S-DNA. More detailed computer simulations~\cite{entropy2} are required to investigate this issue. Instead we ask what are the likely kinetic consequences of assuming that overstretching can involve two double-stranded forms of DNA. This assumption has been made by several authors~\cite{bustamante, cocco,s_form1, s_form2, s_form3,zhou}. We extend our model accordingly.

In `B-to-S' mode we permit basepairs to adopt the forms B, M and U as previously, and in addition permit the adoption of a fourth state, S. We define the thermodynamic properties of this state via the Hamiltonian $H_{\rm S} = \sum_{i=1}^N \cal{H}_i^{\rm S}$, where
\bea
\label{hamil_s}
\cal{H}_i^{\rm S}&=&-\left[\lambda_{\type_i} \e_{\type_i} +w_{\rm S}(f)  \right] \delta_i^{\rm S} \nonumber \\
&+&\frac{1}{2} \eone \, \sum_{\gamma={\rm M}, {\rm U}} \left( \delta_i^{\rm S} \delta_{i+1}^{\gamma}  +\delta_i^{\gamma} \delta_{i+1}^{\rm S} \right) \nonumber \\
&+&\frac{1}{2} \etwo \,   \left( \delta_i^{\rm S} \delta_{i+1}^{\rm B}  +\delta_i^{\rm B} \delta_{i+1}^{\rm S} \right).
\eea
The first line of Equation~(\ref{hamil_s}) describes the extensive free energy of S-DNA. The second term in the first line of Equation~(\ref{hamil_s}), $w_{\rm S}(f)$, is the elastic free energy of S-DNA.  Based on the idea that S-DNA is double stranded and stiff~\cite{cocco}, we assume that it can be parameterized by the wormlike chain in a similar fashion to B-DNA. We assume that $x_{\rm S}(f)$ has a form similar to Equation~(\ref{bform}), with subscripts `B' replaced by subscripts `S'. We assume a contour length $x_{\rm S}=0.58$ nm/bp, and a stiffness constant $S_{\rm S} = 2730$ pN~\cite{cocco}. Lacking experimental measurements of S-DNA flexibility, we assume its persistence length to be equal to the persistence length of B-DNA. The emergent dynamics of our model is qualitatively insensitive to variation of the latter two parameters, which determine the shape (but not the overall scale) of the equilibrium extension curve for S-DNA.
 
The factor $\lambda_{\type_i} $, where $\type_i \in \{{\rm AT}, {\rm CG} \}$, quantifies the diminished base-pairing-stacking energy of S-DNA relative to B-DNA. We determine this loss by requiring phase coexistence between the B and S phases at a pulling force of $f_{\rm o} \equiv 65$ pN, i.e. we fix $\lambda_{\type_i}$ by setting
\beq
\lambda_{\type_i} \e_{\type_i} +w_{\rm S}(f_{\rm o})= \e_{\type_i} +w_{\rm B}(f_{\rm o}).
\eeq 
We obtain $\lambda_{\rm AT} \approx 0.74$ and $\lambda_{\rm CG} \approx 0.77$. We thereby assume that the B-to-S coexistence forces for AT and CG sections are equal. It is possible that this equality does not hold for real DNA~\cite{bustamante,bloomfield1}, but conclusive data do not yet exist. Indeed, there is evidence suggesting that $\lambda$-DNA (50:50 AT:CG) and CG-rich DNA both overstretch at 65 pN~\cite{cs}. The qualitative kinetic features we shall describe are robust to variations in these forces of tens of piconewtons.

The second line of Equation~(\ref{hamil_s}) describes the energetic penalty for interfaces between S-DNA and molten DNA. We assume this to be identical to the B-M junction penalty (and to the B-U and S-U penalties).

The third line of Equation~(\ref{hamil_s}) describes the energetic penalty for interfaces between S-DNA and B-DNA. The parameter $\etwo$ controls the slope (`cooperativity') of the overstretching (B-to-S) plateau at low temperature. In Ref.~\cite{cocco} this was set to $3 \kb T$. We found that this value, in concert with our wormlike chain parameterization for S-DNA,  gave an overstretching transition noticably less cooperative than that observed in experiment (in Ref.~\cite{cocco} a different elastic behavior was assumed for S-DNA). We choose instead to set the junction energy to $\etwo=5\, \kt$, in order to give an overstretching cooperativity similar to that observed in experiment. This choice is without theoretical foundation, and should be regarded as a fit to experimental data. However, the kinetic features we observe are insensitive to variation of this parameter over a range of tens of $\kb T$; the role of $\etwo$ is to set the slope of the B-to-S plateau.

\section{Model dynamics}
\label{secdynam}
With the thermodynamics of our model determined, we investigate its predictions for the dynamics of overstretching. We use a Monte Carlo (MC) dynamics designed to mimic pulling at a constant speed, appropriate for experiments carried out using optical tweezers. 

We consider an $N$ basepair stretch of model $\lambda$-phage DNA. We choose $N$ to be either $4.1 \times 10^4$ (when modeling the experiments of Ref.~\cite{hanbin}) or $4.5 \times 10^3$ (when modeling experiments presented in Refs.~\cite{rief,cs}). We describe its state at a given time $t$ by the vector $\left\{ \alpha_1(t), \ldots,\alpha_i(t),\ldots,\alpha_N(t) \right\}$, where $\alpha_i(t)$ denotes the state $\alpha \in \{$B,M,U,S$\}$ of basepair $i$. To this we attach a Hookean spring designed to model the optical trap. We require that the system (molecule plus trap) be in mechanical equilibrium at all times, i.e. that the tension of the trap and at all points on the molecule have a uniform value, $f$. We thereby assume that mechanical transduction of force (including propagation of
torque) is much more rapid than molecular changes of state. We write the molecular extension as $\ell(f,t)=\sum_{\alpha} n_{\alpha}(t) \,  x_{\alpha}(f)$, where $n_{\alpha}(t)$ denotes the number of basepairs in state $\alpha$ at time $t$, and $x_{\alpha}(f)$ is the extension per basepair of state $\alpha$ as a function of force. The optical trap bead is moved $f/k_{\rm t}$ from the trap centre, where $k_{\rm t}= 0.1$ pN/nm is the trap force constant.

In optical trap experiments the {\em total} extension of the molecule and trap, $L(t)$, is increased at constant speed, $L(t) = v_0 t$. In our simulations we impose a total length of this nature, and calculate the resulting tension $f$ and molecular extension $\ell(f,t)$ by requiring that the imposed extension equal the extensions of the molecule plus trap, $L(t) = \ell(f,t) +f/k_{\rm t}$. The relative extensions of the molecular states are fixed by the constraint of mechanical stability, $f=f_{\rm B}(x_{\rm B})=f_{\rm U}(x_{\rm U})=2f_{\rm M}(x_{\rm M})=f_{\rm S}(x_{\rm S})$, where the $f_{\alpha}(x)$ are found by inverting the functions $x_{\alpha}(f)$. 

In pulling simulations we impose fixed boundary conditions, with sites 1 and $N$ constrained to be B-form (modeling `clamped' ends). We start each run from a fully helical configuration at imposed length $L_0=0.31 N$ nm, which implies $N x_{\rm B}(t=0)+f_{\rm B}(x_{\rm B})/k_{\rm t}=L_0$. This relation determines the initial molecular tension $f(t=0)$ and the initial extensions $x_{\alpha}(t=0)$. We equilibrate the system at this fixed, small extension before beginning the pulling protocol. To perform a pulling simulation we increment the global length according to $L(t) = v_0 t$, with $t$ measured in units of Monte Carlo sweeps (on average one update per site). When the force exceeds 120pN (600 pN in Figure~\ref{figbstwo}), we reverse the loading rate ($v_0 \to - v_0$). A Monte Carlo step consists of a proposed change in the state $\alpha_i \to \alpha_i'$ of a randomly chosen site $i$, accepted with the Glauber probability
\bea
p_{\rm acc}(\alpha_i \to \alpha_i')=\left[1+ \exp(\beta \Delta \cal{H}_i) \right]^{-1}.
\eea
Here $\beta \equiv \left(\kb T\right)^{-1}$, and
\bea
\Delta \cal{H}_i &\equiv& \cal{H}\left[ \left\{\alpha_1, \ldots,\alpha_i',\ldots,\alpha_N \right\} \right] \nonumber \\
&-&\cal{H}\left[ \left\{\alpha_1, \ldots,\alpha_i, \ldots,\alpha_N \right\} \right]
\eea
is the change in Hamiltonian $\cal{H}$ upon the change in state of basepair $i$ from $\alpha_i$ to $\alpha_i'$. The thermodynamic parameters of our model enter the rates for changes in basepair state via the Hamiltonian $\cal{H}$, Equation~(\ref{hamiltonian}). Recall that the Hamiltonian is nonlocal by virtue of the loop entropy term for ssDNA. Because Glauber transition rates satisfy detailed balance~\cite{det_bal}, our simulated DNA molecule will evolve toward its equilibrium state at a given force $f$. Whether it reaches this equilibrium state depends on the ratio of rates for changes in molecular state and pulling: very rapid pulling gives the molecule little chance to equilibrate at a given tension. We assume that the fundamental timescale on which individual bases attempt to change state (`basepair closing time') is $\tau_{\rm f} = 28 \mu$s~\cite{bubbles1}, chosen to agree with basepair fluctuation timescales measured using fluorescence correlation spectroscopy~\cite{bubbles0,bubbles1}. This is then the timescale for a single Monte Carlo sweep. We vary the pulling rate over the range 150 to 4000 nm/s, appropriate for experiments reported in Refs.~\cite{hanbin,rief,cs}.

The kinetic data we present are qualitatively robust to a change in fundamental timescale of an order of magnitude. Timescales derived from FCS data associated with the closing of individual basepairs range from $\tau_{\rm f} \sim 10^{-4}$ to $\tau_{\rm f} \sim 10^{-5}$ s~\cite{bubbles0}; our results are qualitiatively unchanged if we adopt as our timescale any value within this range. We find that within our model the fundamental timescale of Ref~\cite{bubbles1} ($\tau_{\rm f} = 28 \mu$s) gives good agreement with an experimentally observed rate-dependent unpeeling transition~\cite{rief,cs} (see Figure~\ref{figbstwo}). 

The question as to what constitutes the `best' timescale for coarse-grained models such as the one we study here is an open question. Nuclear magnetic resonance (NMR) data imply that molecular fluctuations occur on a much shorter timescale, $10^{-7}$ to $10^{-8}$ s~\cite{cocco,bubbles0}, than those associated with FCS data. However, the authors of Ref.~\cite{bubbles0} suggested that NMR fluctuation timescales might be sensitive to fluctuations smaller than those necessary to effect the opening or closing of a single basepair. Furthermore, a simple kinetic model employing a basepair closing time of 28 $\mu$s was shown in Ref.~\cite{bubbles1} to give reasonable agreement with experimental FCS data.

In Ref.~\cite{cocco} a base-closing timescale of $10^{-8}$ s was adopted and used to derive, for a simple kinetic model, predictions for the nature of the hysteresis seen during DNA overstretching. The model concerned the advance and retreat during overstretching of the unpeeled `front' (the junction between unpeeled and hybridized DNA) over an inhomogeneous sequence. The authors showed that considerable `roughness' in force-extension data arose during shortening as a consequence of the `stick-slip' motion of the front as it traversed the inhomogeneous sequence. Although the timescale chosen in that work is seemingly at odds with our choice of a much slower FCS-derived timescale, we believe the apparent inconsistency can be explained by the difference in dynamical procedures used in our paper and in Ref.~\cite{cocco}. In this paper we use a heat bath (Glauber) dynamics in which thermal fluctuations permit the system to explore its free energy landscape. As a consequence, the emergent dynamics of our model incorporates processes that occur on a broad range of timescales, including those that require the crossing of large free energy barriers (called `activated processes'). By contrast, the dissipative dynamics employed in Ref.~\cite{cocco} permits no processes involving thermal fluctuations, instead corresponding to a monotonic descent to the local free energy minimum. Thermal processes are considered in an approximate fashion by smoothing (averaging) the local sequence. In the context of the model discussed in~\cite{cocco}, this dynamics amounts (for a given global extension) to relaxation of the unpeeled front toward the local free energy minimum of the sequence. As total extension is incremented, this local minimum is changed, and the unpeeled front may relax again to the new local minimum. We demonstrate in Appendix B that this dissipative dynamics is largely insensitive to the fundamental timescale employed, provided that the unpeeled front has time to reach the local free energy minimum. We show that this is so for a given set of conditions for timescales ranging from $\tau_{\rm f} = 2.8 \times 10^{-5}$ s to $\tau_{\rm f} = 10^{-8}$ s. We therefore believe there to be no direct contradiction between our chosen Monte Carlo timescale of 28 $\mu$s and the $10^{-8}$ s adopted in Ref.~\cite{cocco} as the timescale for a dissipative dynamics. The difference in fundamental timescales is not evident in force-extension data because of the different dynamical protocols used. In the context of our model, we find the FCS timescales to give much better agreement with experiment when considering the slow, high-force unpeeling transition following the initial overstretching plateau (Refs.~\cite{rief,cs} and Figure~\ref{figbstwo}).

In parallel with proposed changes of molecular state we propose $N$ times per MC sweep a change $\delta f$ in the global tension, accepted if the quantity $|\sum_{\alpha} n_{\alpha}  x_{\alpha}+f/k_{\rm t}-L(t)|$ does not increase. This frequent adjustment of force enforces the condition of mechanical equilibrium.

The behavior of U-DNA is governed by the following rules. We allow unpeeling to start only from nicks, which we place randomly in every simulation (in some simulations we consider un-nicked DNA). We consider all nicks to be on the same strand, ensuring that one strand is always load-bearing. A nick is made between randomly-chosen neighbouring sites, $i_{\rm n}$ and $i_{\rm n}+1$. Unpeeling can proceed left from $i_{\rm n}$ and right from $i_{\rm n}+1$. Any unhybridized site connected to at least one nicked site by an unbroken chain of unhybridized sites is defined to be unpeeled; if not so connected, it is defined to be molten. The effect of local changes of state associated with a change in state of a single site may be non-local. For example, annealing an unpeeled strand of length $\ell_1$ at a position $\ell_2<\ell_1$ from its end changes a region of length $\ell_1-\ell_2$ from unpeeled to molten. Such a process is generally very unfavourable, however, being suppressed by the thermodynamic predominance of U- over M-DNA. If unpeeled and molten regions merge, the latter is subsumed by the former (as is appropriate given our definitions of `unpeeled' and `molten'). If an unpeeled front advances from its starting nick to an adjacent nick, the corresponding unpeeled region is considered to be completely detached, and is not permitted to change state for the remainder of the simulation.
 
Kinetic data shown in the following sections are representative individual trajectories, selected at random from a large number of such trajectories. On some plots we show symbols denoting equilibrium force-extension values. Each equilibrium value was calculated by averaging four independent simulations run at zero loading rate; error bars are smaller than symbols. In Figures~\ref{figfmone} and~\ref{figfmthree} we show in addition equilibrium force-extension data for nicked systems in force-melting mode. Because in this case complete enumeration of microstates is straightforward, these data were calculated by evaluating numerically the partition function using the Hamiltonian of our model and the sequence of $\lambda$-DNA, assuming a nick between basepairs 2 and 3.

\section{Simulation results in the presence of nicks}
\label{sec_nicks}

\subsection{B-to-S mode}
\label{sec_s2}
\begin{figure*}[tb] 
\centering
\includegraphics[width=3.2in]{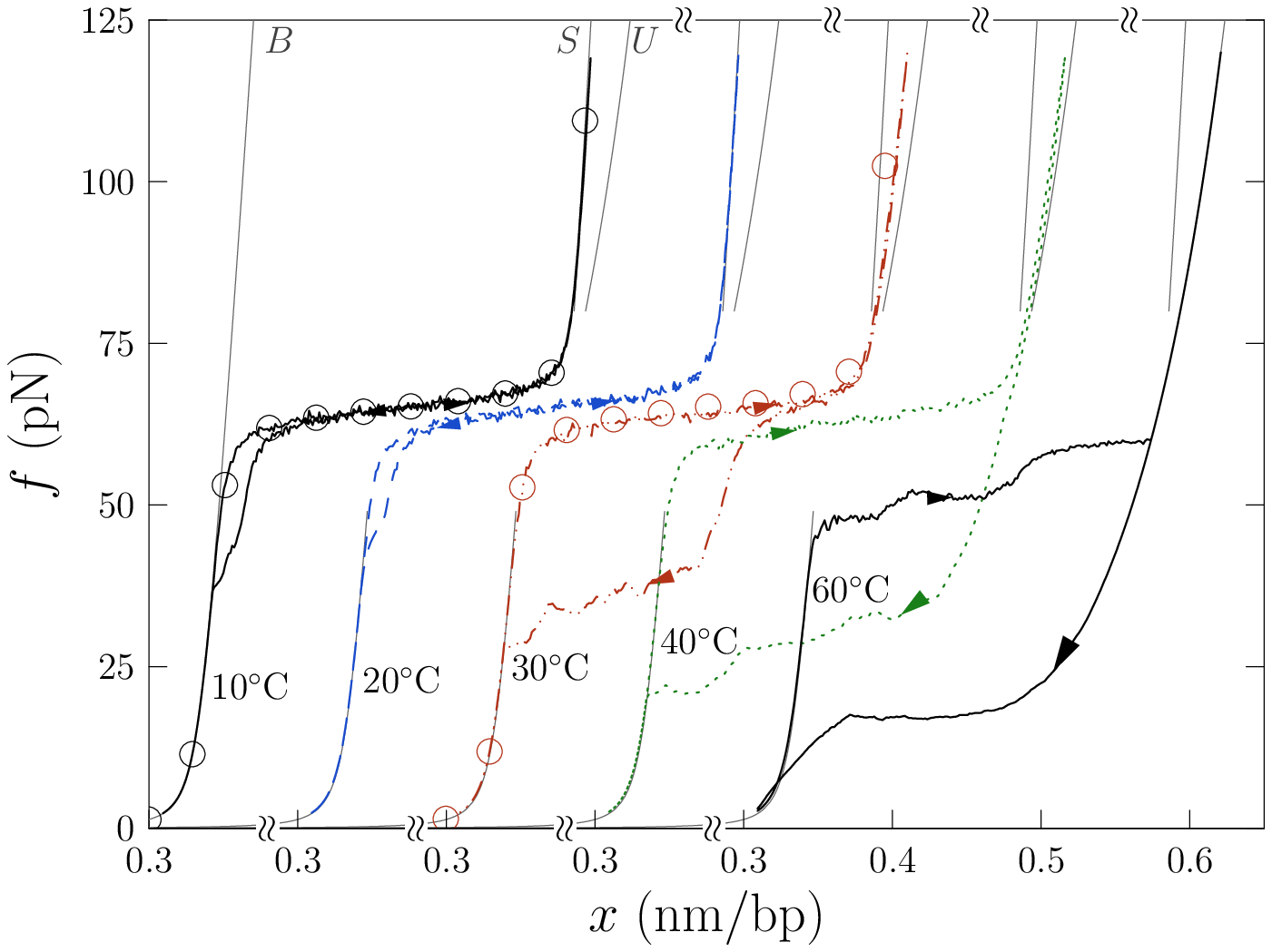} 
\includegraphics[width=3.7in]{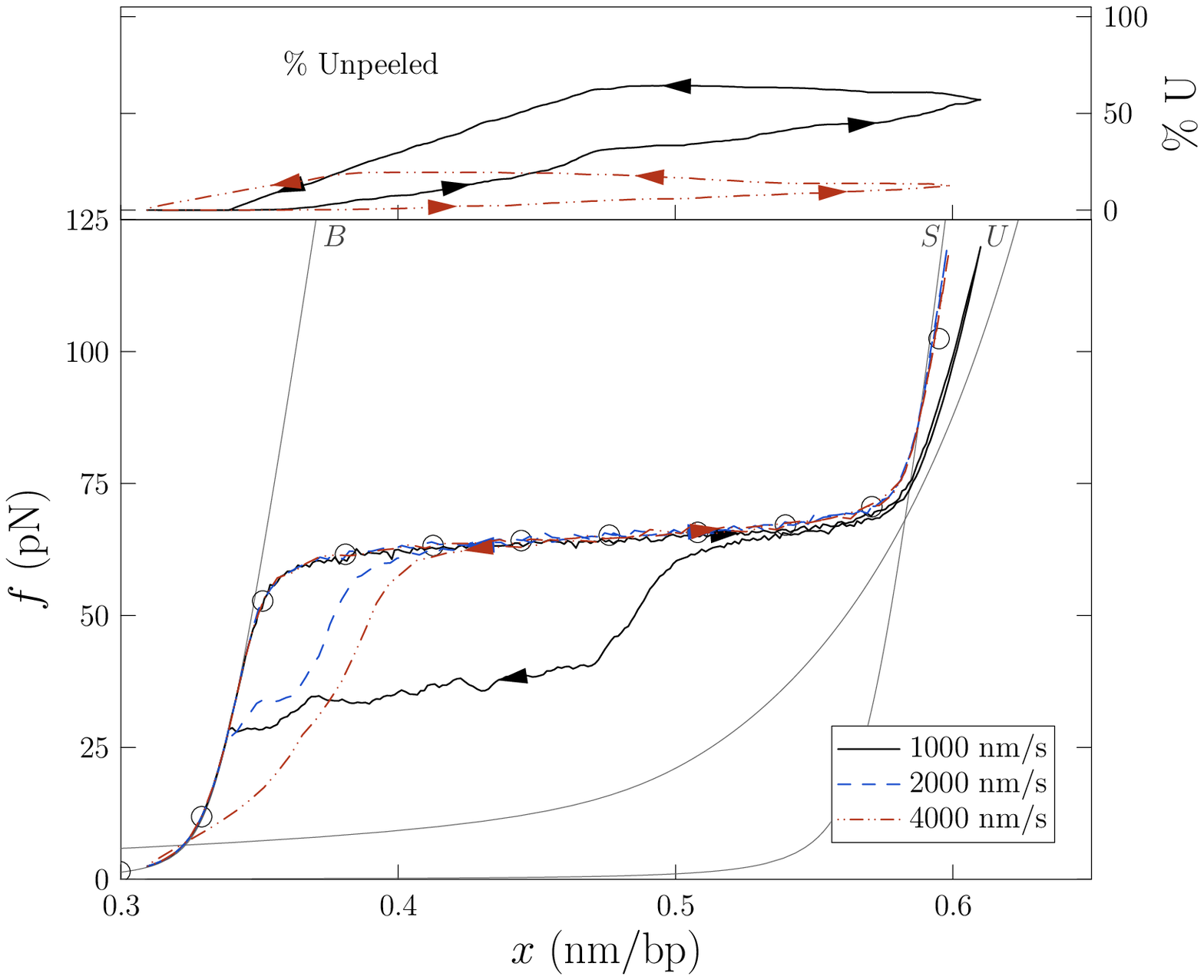} 
\caption{\label{figbsone} B-to-S mode: Effect of temperature. Simulated force-extension data for once-nicked $\lambda$-phage DNA at $500$ mM NaCl, chosen to model experiments presented in Ref.~\cite{hanbin}. {\bf Left panel.} We show data for pulling rate $1000$ nm/s at $T=$ 10, 20, 30, 40 and 60$\dc$ (offset left to right), together with equilibrium data for 10 and 30$\dc$ (circles). At low temperature overstretching is a transition from B- to S-DNA. Increasing temperature causes unpeeling, resulting in the onset of `asymmetric' hysteresis and a drop in plateau height. The progressive increase with temperature of the degree of hysteresis results from a competition between the entropy of U-DNA and the basepairing energy of S-DNA. The overstretched state at high temperature is largely U-DNA. The grey lines here (and in subsequent figures) are equilibrium force-extension data for pure states. {\bf Right panel.} For 30$\dc$ we show data obtained at pulling rates of 1000, 2000 and 4000 nm/s, together with the percentage of unpeeled DNA during stretching and shortening at the slowest and fastest rates (top panel; arrows and line colors correspond to those on force-extension curves). The slow initial unpeeling is largely disguised by the similarity of the S and U force-extension relations in this regime (see, however, Figure~\ref{figbstwo}). By contrast, the slow annealing is made evident by the distinct nature of the U and B force-extension relations at low forces. In this example increasing the pulling rate {\em reduces} the degree of hysteresis observed, because one strand has less time to detach from the other.}
 \end{figure*}
 
We observe a rich dynamics in B-to-S mode, due to the interplay of two hybridized states (B and S) and, chiefly, one unhybridized state (U). In Figure~\ref{figbsone} (left panel) we show the temperature dependence of the overstretching behavior of once-nicked $\lambda$-DNA at $500$ mM NaCl and pulling rate $1000$ nm/s, chosen to model the experiments presented in Ref~\cite{hanbin}. We set the loop exponent $c$ to 2.1~\cite{footnote0}. At low temperature, well below the zero-tension melting point of our model at this salt concentration $(T_{\rm m} \approx 87\dc)$, the hybridized state is stable with respect to the unhybridized state over the full range of force. As determined by the thermodynamics we have imposed, we observe a transition from a B-rich to an S-rich phase at the B-to-S overstretching force of $65$ pN. Because the interconversion of two hybridized states involves only local, and therefore modest, free energy barriers, we observe a reversible dynamics associated with the B to S transition.
\begin{figure*}[ht] 
\centering
\includegraphics[width=3.7in]{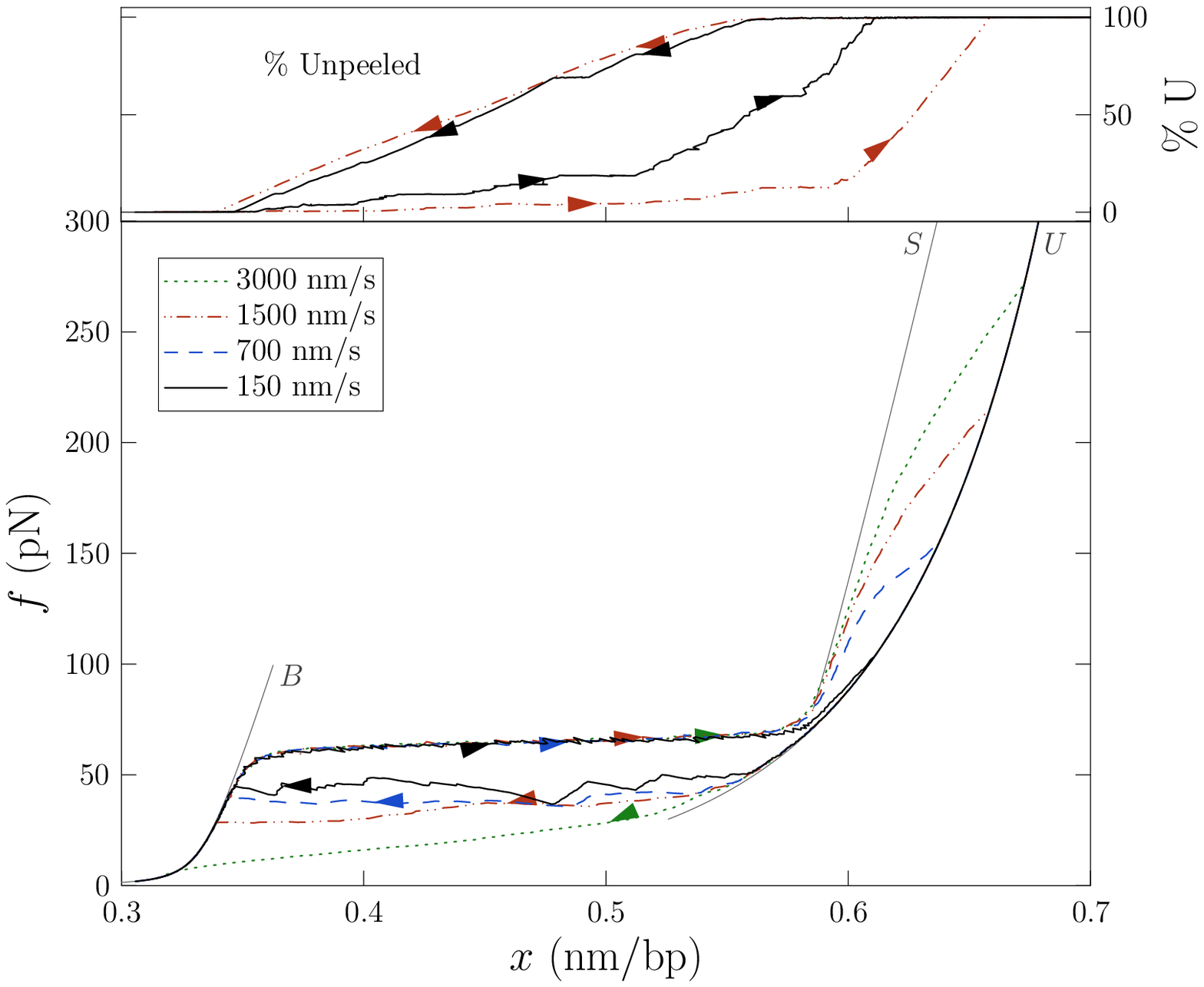} 
\includegraphics[width=3.2in]{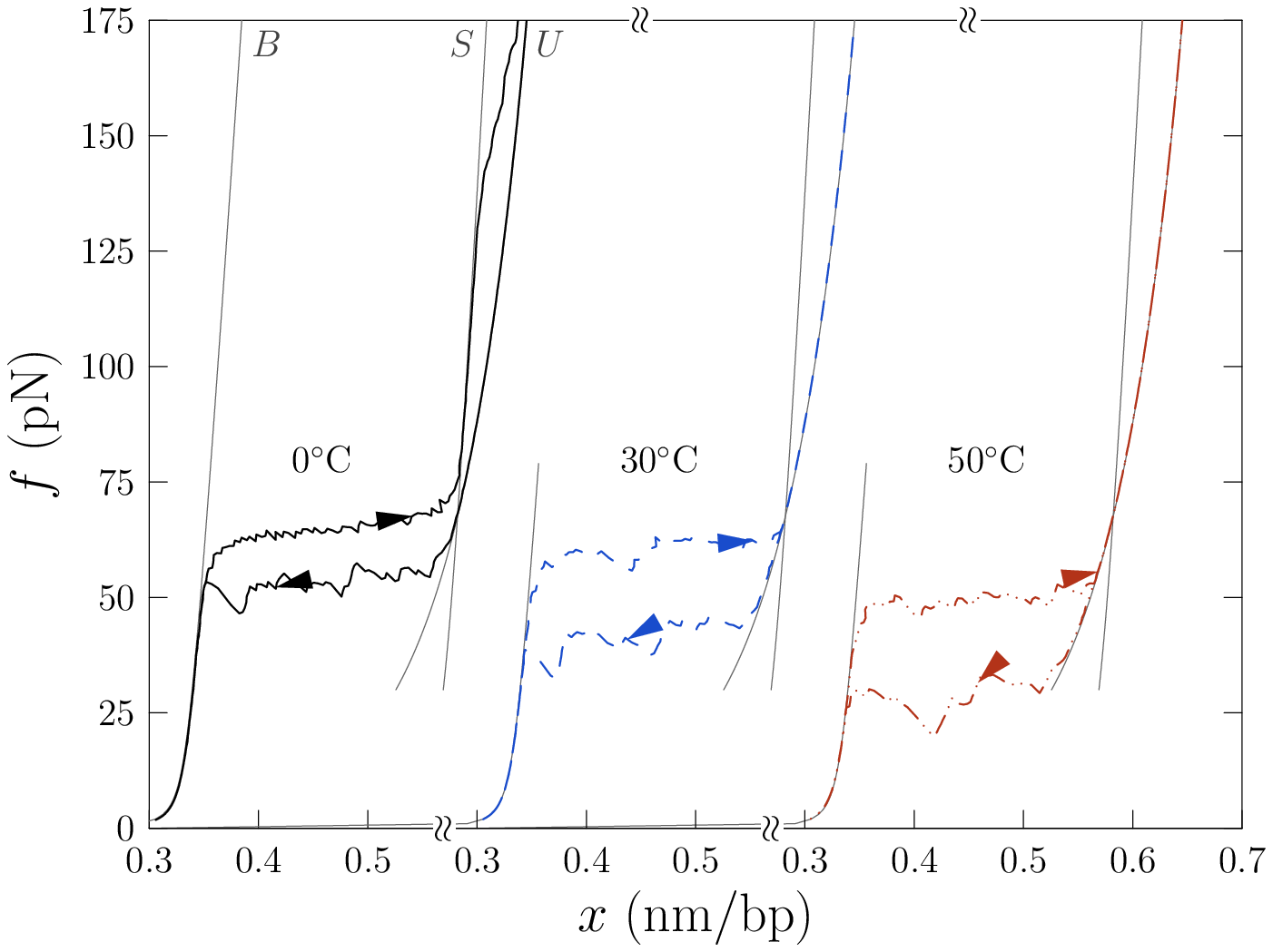} 
\caption{\label{figbstwo} B-to-S mode: high-force transition. Simulated force-extension data for a once-nicked 4.5 kbp fragment of $\lambda$-phage DNA at $150$ mM NaCl, chosen to model experiments reported in Refs.~\cite{rief,cs}. {\bf Left panel.} Effect of pulling rate at $20 \dc$. We employ pulling rates of 3000, 1500, 700 and 150 nm/s. Following the B-to-S transition at 65 pN, we observe a strongly rate-dependent unpeeling transition at forces in good agreement with data presented in Refs.~\cite{rief,cs}. We infer from this comparison that the correct fundamental timescale for our coarse-grained model lies in the 10-100 $\mu$s range characteristic of fluorescence correlation spectroscopy data~\cite{bubbles0,bubbles1}. We show also the fraction of unpeeled DNA for pulling rates 150 and 1500 nm/s (top panel; arrows and colors correspond to those of force-extension data). {\bf Right panel.} Effect of temperature at pulling rate 150 nm/s. We show data for temperatures 0, 30 and 50$\dc$. At sufficiently high temperature the unpeeling transition pre-empts the B-to-S conversion, lowering the overstretching plateau.}
 \end{figure*}

 As temperature increases, partial unpeeling begins to occur. U-DNA is longer than B-DNA at tensions $\geq 10$ pN, and so becomes increasingly favoured thermodynamically as tension increases. In this regime we observe the progressive increase with temperature of the degree of an hysteresis that is asymmetric in character: the overstretching force remains pulling rate-independent for moderate temperatures, whereas the shortening behavior is strongly rate-dependent. The degree to which we observe hysteresis at a given temperature is similar to that seen experimentally in Ref~\cite{hanbin}. In our model, this hysteresis arises from the slow dynamics induced by the macroscopic displacement of the nicked front as it moves basepair-by-basepair. The origin of the hysteresis {\em asymmetry} is the interplay between the phases B, S and U, combined with the mechanical behavior of each. We clarify this point in Figure~\ref{figbsone} (right panel) by showing for 30$\dc$ the percentage of the molecule in unpeeled form. At this temperature the initial transition is B-to-S, occurring in equilibrium at 65 pN. We also observe a slow unpeeling (conversion of B to U and later S to U) slightly in arrears of the B-to-S transition. However, this unpeeling is largely masked in force-extension data by the similarity of the S and U force-extension relations in the relevant regime. By contrast, the slow annealing (conversion of U to B) observed during shortening is made evident by the distinct nature of the U and B force-extension relations at low forces.
 
As temperature increases further, unpeeling causes the overstretching force to drop (above 30$\dc$ in Figure~\ref{figbsone}). As hybridized DNA becomes less stable with respect to unhybridized DNA, force-induced unpeeling (and hence molecular elongation) can occur at tensions at or below 65 pN, concurrent with or pre-empting the B-to-S transition. In this event the plateau height falls, and becomes rate-dependent. When considerable unpeeling is observed the plateau also `roughens', fragmenting into successive mini-plateaux (left panel of Figure~\ref{figbsone}, data for 60 $\dc$). This effect reflects the slow, punctuated dynamics of the unpeeled front as it crosses an heterogeneous sequence.

The degree of unpeeling-driven hysteresis in B-to-S mode depends not only on temperature, but also on the position of the nick and on the pulling rate. If the nick, which we place randomly in every simulation, occurs in a CG-rich region (the first~23 kbp of $\lambda$-DNA are roughly 15$\%$ richer in CG basepairs than in AT basepairs), then unpeeling faces a greater barrier to proliferation than had the nick been placed in an AT-rich region. Thus in Figure~\ref{figbsone} while hysteresis in general increases with temperature, the effect of random nick placement can perturb this trend (see results for 10$\dc$ and $20\dc$).  Hysteresis is in addition a non-monotonic function of the pulling rate. At very low pulling rates the molecule can remain in equilibrium. At a pulling rate of 1000 nm/s, hysteresis can be appreciable (see result for 30$\dc$, Figure~\ref{figbsone}). At higher rates (2000 and 4000 nm/s; see Figure~\ref{figbsone}, right panel), the degree of hysteresis {\em decreases}, because unpeeling has less time to proliferate. Again, the effect of random nick placement can disrupt this trend.

More than one transition may be observed in B-to-S mode. Motivated by experiments reported in Refs~\cite{rief,cs}, we perform pulling simulations at salt concentrations of 150 mM up to a maximum force of 600 pN. Because at high forces ssDNA is more easily extended than our model of S-DNA, we observe for a range of temperature a nonequilibrium unpeeling transition at very high forces (upwards of about 150 pN), following the 65 pN equilibrium B-to-S transition. A similar interpretation was given to the two-stage overstretching in Refs.~\cite{rief,cs}. We show representative data in Figure~\ref{figbstwo} (left panel). The second transition (S-to-U) occurs out of equilibrium, as evidenced by the pulling rate-dependence of the transition force. If the terminal pulling force is large enough that this transition is observed, the stretching-shortening cycle becomes hysteretic. This is true even at temperatures sufficiently low that the B-to-S transition shows no hysteresis when the pulling is reversed before the force exceeds $\sim$150 pN. In Figure~\ref{figbstwo} (right panel) we show that for fixed pulling rate the unpeeling transition can at high temperature pre-empt the B-to-S transition, leading to a drop in plateau height. An apparently similar scenario was observed experimentally in Refs.~\cite{rief,cs}, and was discussed theoretically in Ref.~\cite{cocco}. Note that while the high-force unpeeling in our model is strongly rate-dependent, the initial B-to-S transition is not, occurring at 65 pN for all pulling rates from 150 to 3000 nm/s. Similar behavior was seen in experiment~\cite{rief, cs}.

The force at which we observe the S-to-U transition varies strongly with pulling rate, but the values we obtain are similar to those found in experiment at the same pulling rates (e.g. Figure 1 of Reference~\cite{rief}). Assuming that our proposed mechanism for this transition is correct, we infer from this comparison that the basic fluctuation timescale we have adopted (28 $\mu$s~\cite{bubbles1}) is physically appropriate for our coarse-grained model.

In these high-force simulations we assume a na\"ive extrapolation of all elasticity data to large forces. We also dispense with $g(f)$, the force-mediated correction to the loop entropy of ssDNA, because retaining this factor leads to a change in sign of the surface tension of dsDNA-ssDNA junctions at very large forces.

\subsection{Force-melting mode}
\label{secfm}
\begin{figure*}[ht] 
\centering
\includegraphics[width=3.2in]{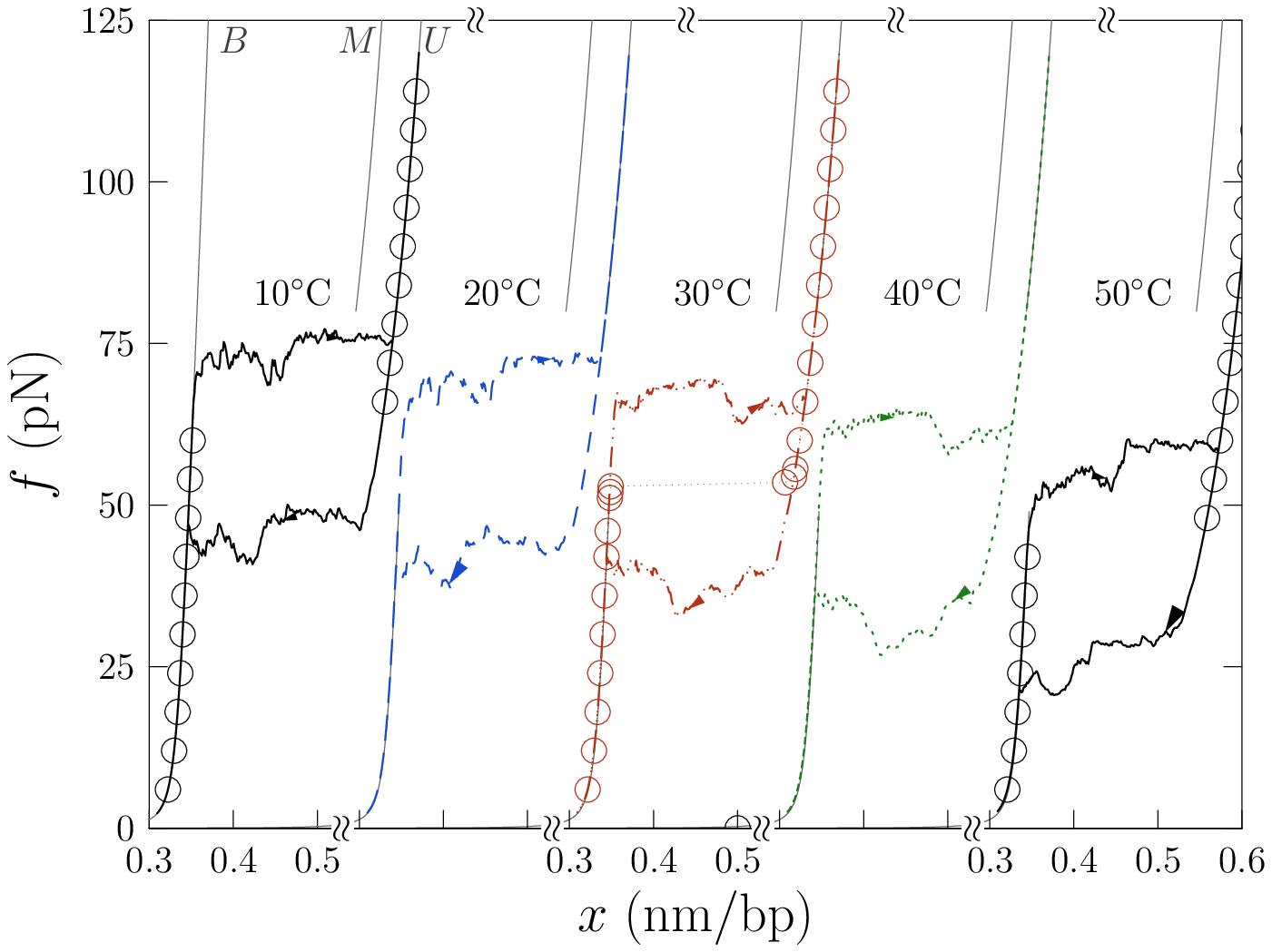} 
\includegraphics[width=3.7in]{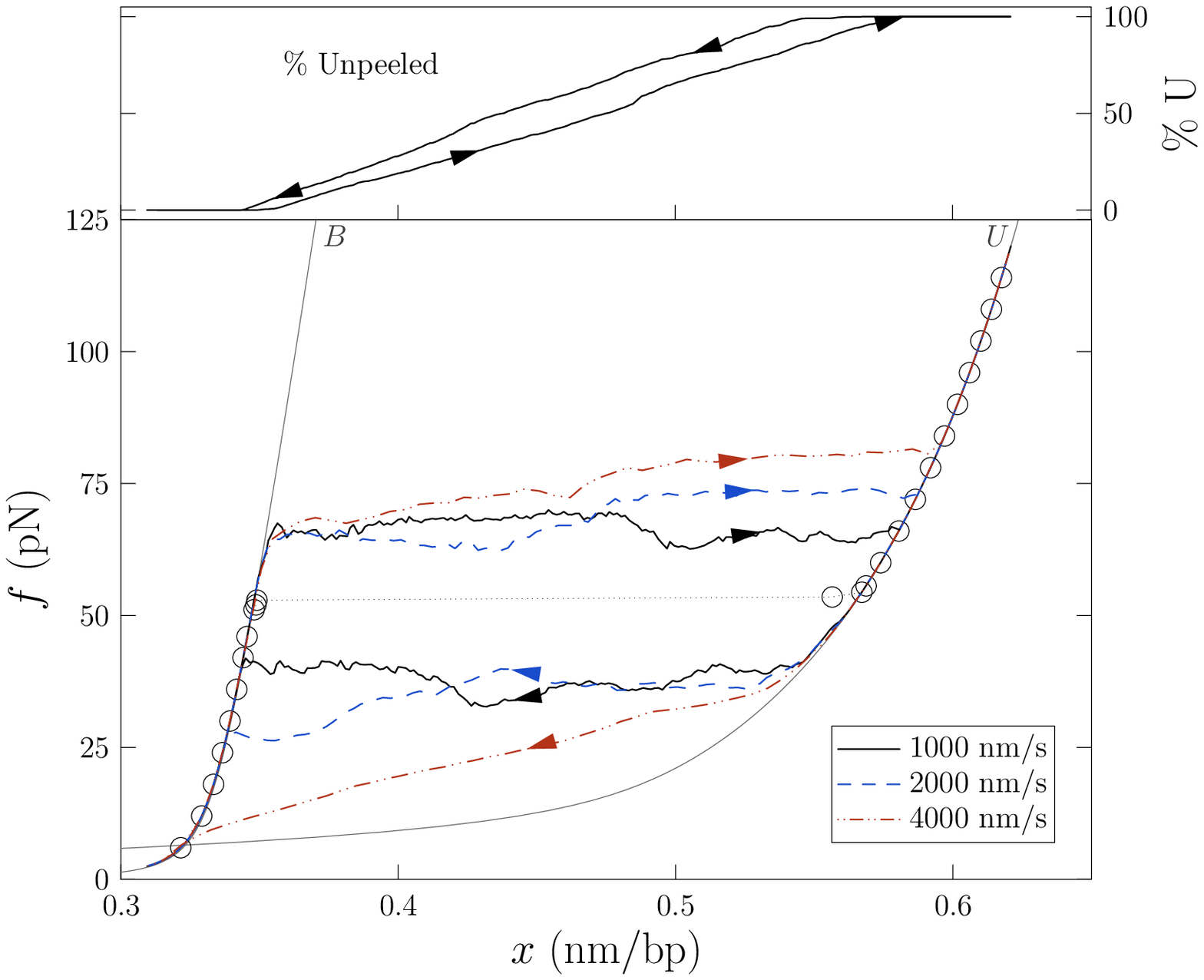} 
\caption{ \label{figfmone} Force-melting mode: Effect of temperature. Simulated force-extension data for once-nicked $\lambda$-phage DNA at $500$ mM NaCl, chosen to model experiments presented in Ref.~\cite{hanbin}. {\bf Left panel.} We show data for pulling rate $1000$ nm/s at $T=$ 10, 20, 30, 40 and 50$\dc$ (offset left to right), together with equilibrium data for 10 and 30$\dc$ (circles). Overstretching corresponds to the conversion of B- to U-DNA, with accompanying hysteresis. This hysteresis is {\em symmetric}, in the sense that the stretching transition lies as far from equilibrium as does the shortening transition. There is no appreciable change with temperature in the degree of hysteresis, because no competition between U- and S-DNA can arise (we do not permit the latter state in force-melting mode). {\bf Right panel.} For 30$\dc$ we show data obtained at pulling rates of 1000, 2000 and 4000 nm/s, together with the percentage of unpeeled DNA during stretching and shortening at 1000 nm/s (top panel; arrows and line color correspond to those on force-extension curve). The slowness of unpeeling confers a rate-dependence upon the overstretching plateau.}
   \end{figure*}
   
In force-melting mode our model can access only states B, M and U, and corresponds to our interpretation of the force-melting theory of DNA overstretching~\cite{bloomfield1,bloomfield2,force_melting, williams, melting_prl}. In Figure~\ref{figfmone} we show data obtained under conditions identical to those used to generate Figure~\ref{figbsone}, chosen to model the experiments reported in Ref.~\cite{hanbin}. As the force on the DNA increases, B-DNA becomes unstable to the unhybridized forms M and U. Stretching results in the complete unpeeling of the DNA. This is so because of the thermodynamic predominance of U- over M-DNA, and also partially because wherever these two phases meet the former subsumes the latter (as is appropriate given our definitions of `molten' and `unpeeled'). Because the dynamics of the unpeeled front is slow, accompanying this unpeeling is considerable hysteresis. However, unlike in the presence of the S-state, unpeeling-driven hysteresis in force-melting mode is {\em symmetric}: the stretching transition lies roughly as far from equilibrium as does the shortening transition. This may be seen clearly by comparing kinetic and equilibrium data in Figure~\ref{figfmone}. This symmetry derives from the fact that strands detach or re-anneal in a similar fashion, namely, via the sequential advance or retreat of the unpeeled `front'~\cite{footnote}: the kinetics of strand detachment is no less sluggish than that of strand re-annealing. The advance and retreat of the unpeeled front occurs by the same mechanism in the presence of S-DNA, but there an asymmetry in force-extension data is induced by the interplay of three phases, B, S and U. 
 
The sluggishness of the unpeeling transition confers upon overstretching data an hysteresis at all temperatures. The symmetry of the hysteresis means that both shortening and stretching occur out of equilibrium, and therefore that the plateau height exhibits a rate-dependence at all temperatures. By contrast, as discussed in the previous subsection, there is in experiment a range of temperature over which the transition is reversible and the overstretching plateau appears to be pulling rate-independent~\cite{hanbin,rief,cs}. In addition, there exists in force-melting mode no analogue of the high-force transition~\cite{rief,cs} observed in B-to-S mode (Figure~\ref{figbstwo}), because the unpeeled state {\em is} the overstretched state: there is no additional phase to which a transition may be made.

In order to prevent complete unpeeling of the DNA in the presence of nicks, we would require within our model a highly collective free-energetic or kinetic `barrier' between molten and unpeeled DNA, or a different notion of what are the `molten' and `unpeeled' states of DNA. We discuss this possibility in Section~\ref{conc}.

\section{Simulation results in the absence of nicks}
\label{sec_nonicks}

\subsection{Force-melting mode}
\label{secfm2}

In this section we examine the predictions of our model in the absence of nicks. Because we forbid fraying from the ends of our simulated DNA molecule (the first and final basepairs are constrained to remain hybridized for the duration of the simulation), unpeeling cannot occur. Nicks allow one strand to rotate freely about the other during stretching. In the absence of nicks, we nonetheless assume that the DNA remains torsionally  unconstrained; if torsionally constrained, the overstretching transition occurs in excess of 100 pN~\cite{cs}. Torsional freedom in the absence of nicks might be engineered by using a careful tethering arrangement.

\begin{figure}[tb] 
   \centering
        \includegraphics[width=3.2in]{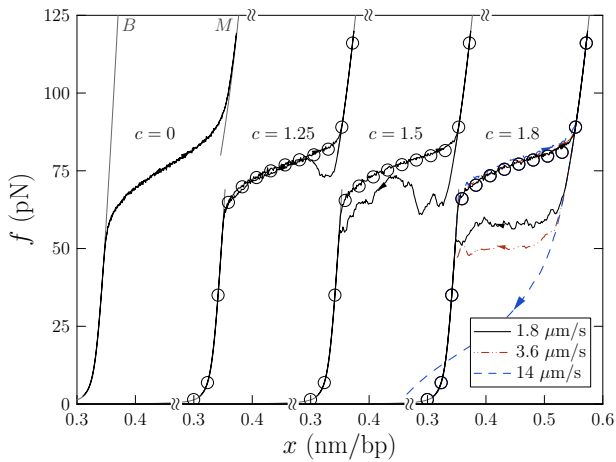} 
   \caption{   \label{figfmtwo} Force-melting mode in the absence of nicks: Effect of varying loop exponent. Simulated force-extension curves for un-nicked $\lambda$-phage DNA at $150$ mM NaCl, $T=25\dc$, and pulling rate 1.8 $\mu$m/s. The curves correspond to loop exponents $c=0$, 1.25, 1.5 and 1.8 (offset horizontally), and we show equilibrium data in three cases (symbols). Hysteresis increases as $c$ increases, even though the {\em equilibrium} fraction of M-DNA decreases with increasing $c$. Note that the `cooperativity' (or sharpness) of the force-melting transition decreases as $c$ decreases. For the largest value of $c$ we show the effect of pulling rate upon the degree of hysteresis.}
   \end{figure}

When unpeeling is suppressed, the only form of unhybridized DNA accessible is the molten state. Consequently, as shown in Figure~\ref{figfmtwo}, we observe in force-melting mode a transition from B-DNA to M-DNA upon stretching. This transition occurs at roughly 70 pN at room temperature, as per the thermodynamic analysis of Refs.~\cite{force_melting, bloomfield1,bloomfield2}. Note that omitting the specific heat correction in Equation~(\ref{entropy}) leads to much larger overstretching forces at low temperature (see Figure~\ref{fignrg})~\cite{rouzina1}. The plateau would be lowered still further by adopting larger values of $\Delta C_p$ still within experimental uncertainty (e.g. $\approx 70$ cal/(mol$\cdot$K$\cdot$bp)). 

When melting is appreciable we observe hysteresis whose extent is strongly dependent upon the value of the loop exponent $c$. In Figure~\ref{figfmtwo} we show data for un-nicked $\lambda$-DNA at fixed temperature (25$\dc$) and salt concentration (150 mM NaCl), for varying loop exponent $c$. While the apparent sharpness of the equilibrium transition (circles) is modified only weakly by variation in the loop exponent (this sharpness is partially counteracted by sequence heterogeneity), the degree of hysteresis is strongly dependent upon this parameter. For values of $c$ appropriate at zero tension ($c \sim 2$) we see considerable hysteresis at typical experimental rates of pulling.
      \begin{figure}[tb] 
   \centering
         \includegraphics[width=3.7in]{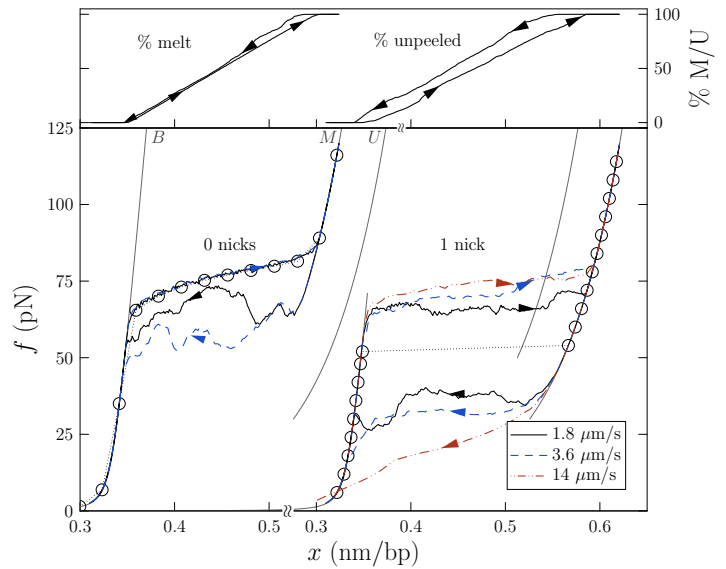} 
   \caption{\label{figfmthree} Illustration of the difference in the character of hysteresis observed in force-melting mode when unpeeling is suppressed and allowed. {\bf Left panel}: un-nicked $\lambda$-DNA, $c=1.8$, $T=25\dc$, 150 mM NaCl, pulling rates 1.8 and 3.6 $\mu$m/s. Melting driven hysteresis is {\em asymmetric} in character, and driven by the loop entropy exponent $c$.   Equilibrium data (symbols) demonstrate that stretching is in equilibrium. For the slower pulling rate we show the percentage of molten DNA during stretching and shortening (top panel; arrows correspond to arrows on force-extension curve). Right set: once-nicked $\lambda$-DNA, $c=1.8$, $T=25\dc$, 150 mM NaCl, pulling rates 1.8, 3.6 and 14 $\mu$m/s. Unpeeling-driven hysteresis is {\em symmetric} in character: stretching and shortening force-extension lines lie equally far from equilibrium (symbols). We observe a pulling rate-dependent overstretching force. For the slowest pulling rate we show the percentage of unpeeled DNA during stretching and shortening (top panel; arrows correspond to arrows on force-extension curve).}
   \end{figure}
   
In our model, this hysteresis is driven by the development of large molten bubbles (see Section~\ref{sectwo}). At sufficiently high tension, molten DNA nucleates and spreads. The nonextensive loop entropy rewards bubble coalescence. During shortening, as tension decreases, B-DNA becomes thermodynamically favoured over M-DNA. However, for large values of $c$ bubbles resist pinching, and the nucleation of B-DNA within M-DNA is suppressed. Large molten bubbles thus close in general from their ends. Because this is a gradual process, requiring a collective relaxation over often large distances, such bubbles can persist far into the shortening transition, preventing the molecule from reaching equilibrium. For smaller values of $c$ bubble pinching is much less costly, and equilibrium is attained more readily~\cite{note}.

We find that melting-driven hysteresis is set largely by $c$, and depends only weakly upon temperature. However, if $c$ were strongly temperature-dependent, as has been suggested~\cite{c_temp}, we would observe an hysteresis that becomes more pronounced with increasing temperature. 

In Figure~\ref{figfmthree} we contrast the character of hysteresis observed in force-melting mode in the absence and presence of nicks.

\subsection{B-to-S mode}
\label{sec_s4}
\begin{figure}[tb] 
   \centering
         \includegraphics[width=3.2in]{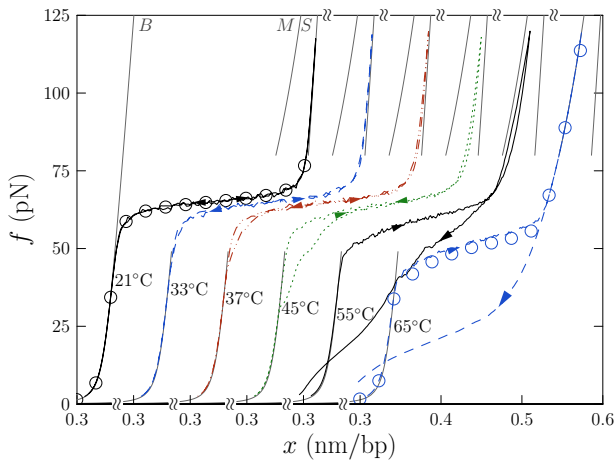} 
   \caption{\label{figbsthree} B-to-S mode in the absence of nicks:  Effect of temperature. Simulated force-extension data for un-nicked $\lambda$-phage DNA at $150$ mM NaCl and pulling rate $3000$ nm/s. We show data for $T=$  21, 33, 37, 45, 55 and 65$\dc$ (offset left to right), with loop exponent $c=1.8$, together with equilibrium data for temperatures $21\dc$ and $45\dc$ (symbols). At low temperature, overstretching is a transition from B- to S-DNA. Increasing temperature causes melting, resulting in the onset of $c$-dependent hysteresis and a drop in plateau height. The overstretched state at high temperature is largely M-DNA.}
 \end{figure}
 
In B-to-S mode, when unpeeling is suppressed, we observe a melting-driven hysteresis and variation in plateau height (Figure~\ref{figbsthree}). These features are also observed in force-melting mode (Section~\ref{secfm2}). However, unlike in force-melting mode, for a fixed value of $c$ this hysteresis is temperature dependent, rendered so by virtue of the temperature-dependent competition between S- and M-DNA. Melting-driven hysteresis sets in at higher temperatures (or lower salt concentrations) than does unpeeling-generated hysteresis, because of the thermodynamic predominance of U- over M-DNA. Note that the data shown in Figure~\ref{figbsthree} were obtained under conditions of 150 mM NaCl (a salt concentration frequently used in experiments); data in the presence of a nick, Figure~\ref{figbsone}, were obtained at a salt concentration of 500 mM (the salt concentration used in the experiments of Ref.~\cite{hanbin}).

\section{Conclusions}
\label{conc}

We have demonstrated that a simple model of the DNA overstretching phenomenon can furnish kinetic predictions on experimental length- and timescales, given as inputs simple experimental thermodynamic and elastic data. This approach offers the possibility of comparing in a detailed manner the two pictures of overstretching. One of our key results is the observation that hysteresis generated by strand unpeeling has a character that depends crucially on the presence of the S-state of DNA, displaying the asymmetry of experimentally-observed hysteresis only when S-DNA is allowed. We observe a progressive increase with temperature in the degree of this hysteresis, owing to a competition between the entropy of U-DNA and the basepairing energy of S-DNA. We have also demonstrated that the nonextensive thermodynamics of ssDNA (loop entropy) is associated with an anomalous kinetics, which may be detectable in overstretching experiments performed upon un-nicked but torsionally unconstrained DNA.

Our interpretation of force-melting theory yields kinetics in generally poor agreement with experiment. The chief cause of this disagreement is the instability of molten DNA to unpeeling: in force-melting mode, our model exhibits complete unpeeling at high forces, conferring upon the overstretching transition a symmetric hysteresis at all temperatures. This instability can be curbed within our model only by interposing between M- and U states a highly collective free energetic or kinetic barrier.

However, we do not discount the possibility that a modified model of force-melting theory might better describe experiment. Based on our work, we pose the following questions. Is there a collective kinetic or thermodynamic effect that prevents unpeeling from nicks in the backbone, rendering melting stable or metastable relative to unpeeling? If so, might a high-force unpeeling follow a lower-force melting, thereby reproducing the two-stage transition shown in Figure~\ref{figbstwo}? Does tension-melted DNA differ fundamentally from thermally-melted DNA (other than in their respective lengths)? Does the loop factor of molten DNA change under tension? Can the loop factor be strongly suppressed without losing much of the extensive entropy of melting? Our interpretation of force-melting theory assumes or results in simple answers to these questions: `no' in each case.

If some or all of these questions could be answered in the positive, we would expect better agreement between model predictions and experimental kinetics. Were melting somehow stable or metastable to unpeeling (perhaps rendered so by hydrodynamic effects or effects of secondary structure) we could in principle observe a high-force unpeeling transition following an initial melting, giving rise to force-extension data analagous to those presented in Figure~\ref{figbstwo}. If the degree of this metastability was temperature-dependent, or if $c$ were temperature-dependent, we might again capture the trend of increasingly prominent hysteresis with increasing temperature, such as shown in Figure~\ref{figbsone}. Our results indicate that a change in $c$ by roughly a factor of 2 over a 40$\dc$ range (i.e. $c \approx 1$ or 1.5 at 10$\dc$ and $c \approx 2$ at $50\dc$) should be sufficient to generate the degree of variation in hysteresis seen in experiment. Interestingly, the variation of $c$ with temperature suggested in Ref.~\cite{c_temp} is consistent with this requirement.

Determining the value of $c$ under force would allow us to make detailed kinetic predictions when large-scale melting proliferates in strained DNA. To measure $c$, one could use the method proposed in Ref.~\cite{c_test}. In this work the authors studied, theoretically, certain autocorrelation functions associated with basepair openings during DNA melting. They demonstrated that these quantities possess a long-time behavior that is a simple function of $c$, and suggested that one could extract $c$ by measuring these correlation functions using fluoresence correlation spectroscopy. We suggest that performing such measurements with the DNA in question held at different tensions may provide a means of discerning the force dependence of $c$.

Measurement of the value of $c$ under tension would allow our model to probe the kinetic behavior of large molten bubbles that develop in strained DNA. For a range of $c$ ($\approx$ 1 to 2) we expect there to be an anomalous kinetics associated with these bubbles. This may have significance for the behavior of the molecule in biological processes. \\

We thank J. Ricardo Arias-Gonzalez and Hanbin Mao for discussions and correspondence, and Carlos Bustamante, Michael F. Hagan, Yariv Kafri, John F. Marko, Felix Ritort, David Sivak and Steven B. Smith for comments or correspondence. We are grateful to Ioulia Rouzina for much valuable correspondence, and for providing a calculation (included in Appendix A) of the tension-dependence of the loop factor of ssDNA.

\section{Appendix A}

{\em Approximate entropy of loops under tension}. The entropy associated with molten DNA has been shown to be important during overstretching from the strong temperature dependence of the overstretching force~\cite{williams}; from theoretical fits to experimental data~\cite{entropy1}; and from molecular dynamics simulations~\cite{entropy2}. Associated with the extensive entropy of melting is a nonextensive penalty for closing a loop, which to a first approximation may be calculated in the following manner~\cite{poland}.

Consider an ideal random walk of $n$ steps on a $d$-dimensional hypercubic lattice, constrained to return to its origin. Because the loop closes, the number of steps in the $+$ and $-$ directions in each dimension must be equal; the number of ways of ordering $m$ such steps in one dimension is $W(m) \equiv m!/[(m/2)!]^2$. Consider a walk in which $r n < n$ steps are taken in a favoured dimension, corresponding to the direction of pulling. The case $r=1/d$ is an unbiased walk. If the remaining $(1-r)n$ steps are distributed equally amongst the other $d-1$ dimensions, then the entropy of the walk is 
\beq
\label{ent1}
S(n,r)= \kb \ln W(r n)+ (d-1) \kb \ln W(\hat{r} n).
\eeq
We have defined $\hat{r} \equiv (1-r)/(d-1)$. The first term in Equation (\ref{ent1}) accounts for the number of ways of ordering the steps in the `pulling' dimension;  the second accounts for the steps in the other $d-1$ dimensions. Equation (\ref{ent1}) can be approximated by Stirling's formula as 
\beq
\label{ent2}
S(n,r) \approx \kb n \ln 2 -\kb c_0 \ln n_r.
\eeq
The first term in Equation (\ref{ent2}) is the extensive entropy accounting for the 2 choices $(+/-)$ for each of the $n$ steps of the walk. The second term is the nonextensive penalty for loop closure. The dependence upon aspect ratio $r$  is confined to the effective loop length parameter $n_r \equiv \frac{1}{2} \pi r^{1/d} \hat{r}^{1-1/d}\, n$. The exponent $c_0 \equiv d/2$ does not depend on $r$. Under zero tension, in $d=3$, this value is increased to $\sim 1.8$ by the constraint of self-avoidance~\cite{fisher}, and further to $\sim 2.1$ if loop interactions are taken into account~\cite{kafri}. 

Stretching indeed changes the nonextensive entropy of a loop. However, in the approximation that we consider here this change in entropy is independent of the length of the loop, and can be regarded as a constant ($n$-independent) correction for all loops. Such is also the case within the freely-jointed chain model (again ignoring excluded volume) in the constant force ensemble~\cite{rouzina1}. This calculation yields the entropic correction in terms of the pulling force $f$:
\beq
\label{ent3}
\Delta S_{\rm loop}(f,r) \approx -\kb c_0 \ln \left(n \right) + \kb \ln g(f),
\eeq
where 
\beq
\label{gee}
\frac{2}{g(f)} =  {\cal L}(f_{\rm r}) \left({\cal L}(f_{\rm r})+\frac{2}{f_{\rm r}^2}-\frac{2}{f_{\rm r}} \coth (f_{\rm r})\right)^{1/2}.
\eeq
Here $f_{\rm r} \equiv f P_{\rm ss }/(\kb T)$ is a reduced force, $P_{\rm ss } \approx 0.7$ nm is the persistence length of ssDNA, and ${\cal L}(x)\equiv 1- [\coth (x) - 1/x]^2 $. The correction $g(f)$ provides an approximate means of accounting for the change in loop entropy with force. Note that $g(f)$ does not tend exactly to unity in the limit $f \to 0$, and so we regard this correction only as a rough estimate of the change in entropy of a loop under pulling. The emergent kinetics of our model does not depend on whether $g(f)$ is present or absent; this term serves only to renormalize the surface tension of dsDNA-ssDNA interfaces (from $\sim 11.5 \kb$T at zero tension to $\sim 6 \kb$T at 65 pN). However, when volume exclusion is accounted for, the correction factor $g(f)$ acquires a dependence upon loop length~\cite{rouzina1}. Instead of adopting this correction explicitly, we choose to vary $c$ in order to probe the regime in which melting is appreciably hysteretic. 

\section{Appendix B}
\label{app2}

{\em What is the appropriate base-closing timescale for our coarse-grained model of DNA?}. As discussed in Section \ref{secdynam}, we use as the fundamental `basepair closing' timescale of our model a value of 28 $\mu$s, derived from fluorescence correlation spectroscopy data~\cite{bubbles0,bubbles1}. In Ref.~\cite{cocco} an NMR-motivated timescale of $10^{-8}$ s was used with a simple model to derive kinetic predictions for the degree and nature of hysteresis of overstretching when unpeeling is permitted. Given that these kinetic predictions appear similar in nature to some of the data presented in this paper, we must explain why this can be so when two dramatically different timescales have been employed. 

We believe that the difference in timescales is not evident because, chiefly, of the different dynamical protocols employed in the two papers. The dissipative dynamics of Ref~\cite{cocco} is not equivalent to the Monte Carlo dynamics we use here. The former describes a dynamics that is dominated by the local free energy landscape of the inhomogeneous sequence, as we show below. While our Monte Carlo dynamics is indeed influenced by DNA sequence, many of the features we describe in this paper, such as the asymmetry of hysteresis in the presence of S-DNA, do not depend solely on sequence heterogenity.

\begin{figure*}[bth!] 
  \centering
\includegraphics[width=3.2in]{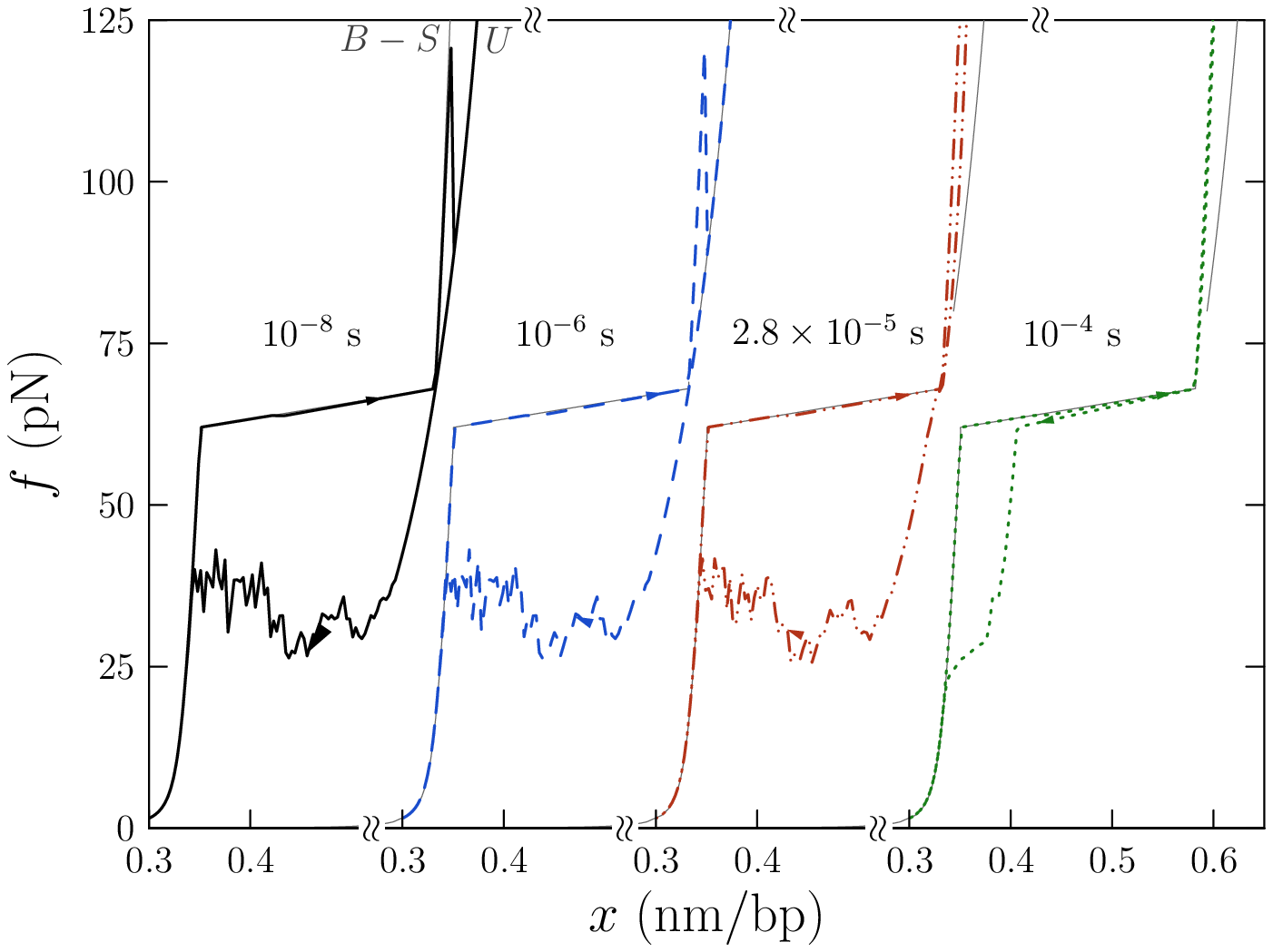} 
\includegraphics[width=3.2in]{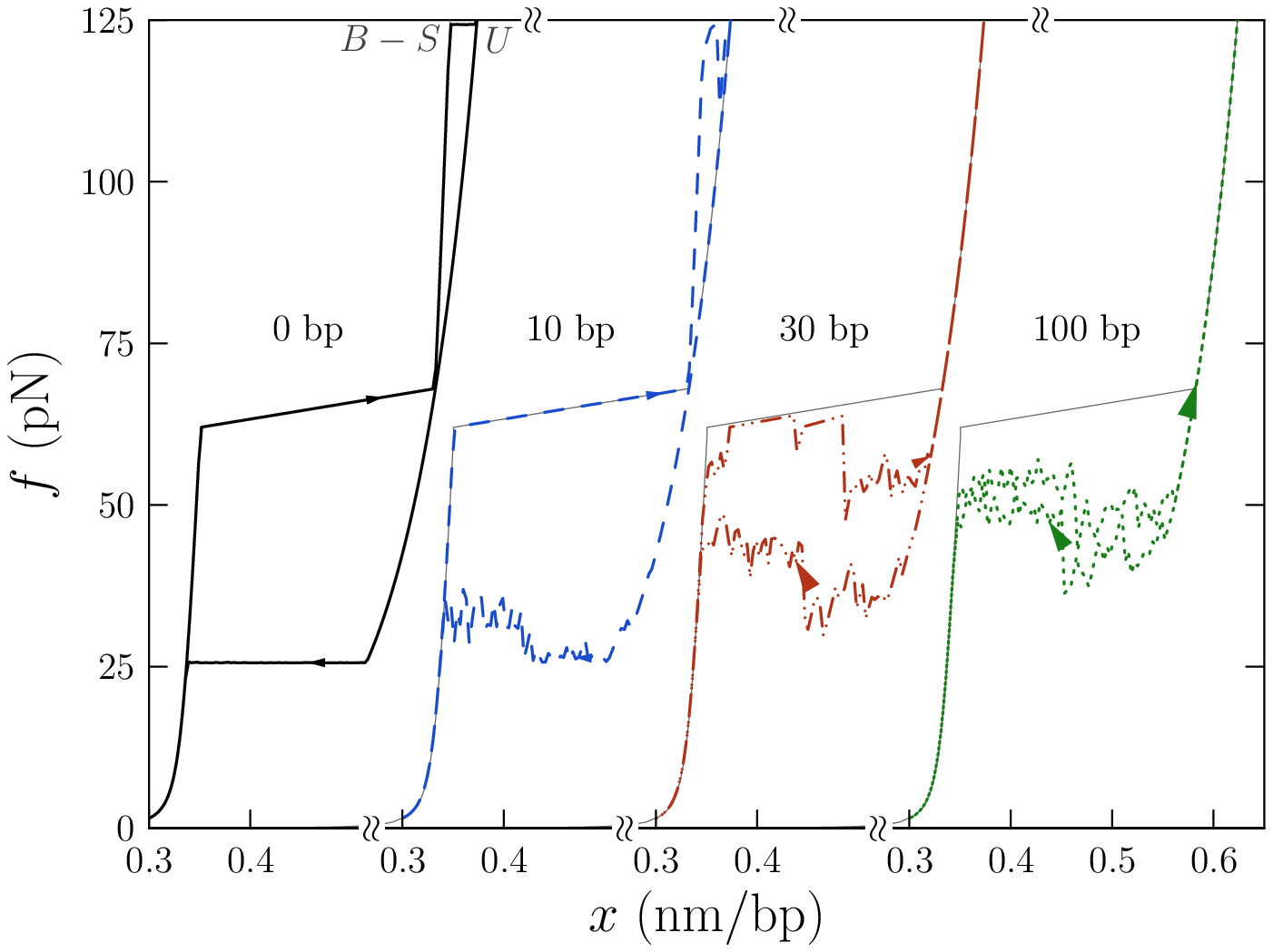} 
 \caption{ \label{figapp} Simulated force-extension data for a simple model incorporating only a hybridized (`B-S') state of DNA and unpeeled (U) DNA, with the dynamical variable being the position of the junction between the two phases (the unpeeled `front'). We show data obtained at 30$\dc$ for 500 mM NaCl, using `zero-temperature' Monte Carlo dynamics (see Appendix B). The pulling rate is 1000 nm/s. {\bf Top panel}: Effect of fundamental timescale upon nature of hysteresis, with sequence averaged over 15 bp. We show data obtained using timescales of $10^{-8}$, $10^{-6}$, $2.5 \times 10^{-5}$ and $10^{-4}$ s. Hysteresis `roughness' does not depend strongly on basic timescale, over several order of magnitude, provided that the unpeeled front has time to reach the local free energy minimum of the averaged sequence. {\bf Bottom panel:}  Effect of sequence-averaging on nature of hysteresis. We show data obtained using a basic timescale of $10^{-6}$ s, with averaging windows of 0 bp (no averaging), 10, 30 and 100 bp. The averaging determines the local free energy minima of the sequence, which influences strongly the dissipative dynamics.}
 \end{figure*}

In Figure~\ref{figapp} we show data obtained from a model closely related to the kinetic model of refence~\cite{cocco}. Our simple model consists of a 41 kbp fragment of $\lambda$-DNA, with each basepair allowed to assume only two states. The first state is a hybrid `B-S' phase, possessing an elastic behavior that mirrors the form of the overstretching plateau (see Ref.~\cite{cocco}), and having the basepairing-stacking energy of the B-DNA of our Monte Carlo model. The second phase is the unpeeled phase, having the thermodynamic and elastic behavior of the U state of our Monte Carlo model. A nick is placed between sites 2 and 3. Starting from the fully hybridized state, we increment the length of the DNA at 1000 nm/s, according to a fundamental timescale $\tau_{\rm f}$, until the tension of the molecule reaches 140 pN. We then reverse the loading rate. We do not model the optical trap.

Because only the `B-S' and U states are accessible, the only relevant dynamical variable is the junction between these phases, which we call the unpeeled `front'. We allow the front to advance or retreat along the inhomogeneous sequence according to a `zero-temperature' dynamics: if a change in position of the front reduces the free energy of the system then this change is accepted; if not, it is rejected. This dynamics is analogous to the dissipative dynamics employed in Ref~\cite{cocco}. We find that the kinetic predictions of this model are dominated by sequence heterogeneity. In Figure~\ref{figapp} (top panel) we demonstrate that the nature (extent and `roughness') of hysteresis does not depend strongly on the fundamental timescale $\tau_{\rm f}$, provided that the nicked front has time to reach the local minimum of the free energy landscape. Whether this is so depends strongly on the nature of the landscape, which is influenced by any `smoothing' that one might apply (in order to account on small lengthscales for thermally-driven behavior~\cite{cocco}). In Figure~\ref{figapp} (bottom panel) we show that averaging the local sequence using windows of 0, 10, 30 and 100 bp influences strongly the zero-temperature dynamics.

The `roughness' of unpeeling-generated hysteresis in our Monte Carlo model is not wholly determined by sequence, but owes its form in part to the response of the optical trap and to thermal noise. Indeed, the `sawteeth' features seen during relaxation differ from run to run; this is not the case for the dissipative dynamics described in this Appendix. Our conclusion from these data is that there is no direct contradiction between the basic timescale we have used in this paper (28 $\mu$s), in concert with Glauber dynamics, and the $10^{-8}$ s employed as the integration step of the dissipative dynamics of Ref.~\cite{cocco}. These different fundamental timescales yield apparently similar predictions for the character of unpeeling-driven hysteresis because predictions are made using different dynamical protocols. Within our model, FCS timescales yield much better agreement with experiment when considering the slow, high-force unpeeling transition following the initial overstretching plateau (Refs.~\cite{rief,cs} and Figure~\ref{figbstwo}).


\begin{thebibliography}{99}
\bibitem{footnote0} The behavior shown in Figures~\ref{figbsone} and~\ref{figbstwo} is essentially insensitive to the choice of $c$, because the form of unhybridized DNA that proliferates is unpeeled, and not molten. Unpeeled DNA possesses no `loop factor'. We ran simulations for values of $c$ from 0 (no molten loop factor) to $c \to \infty$ (melting suppressed), and found no difference in the emergent dynamics reported in these figures.
\bibitem{footnote} We allow reattachment of the two strands at any point along their length. However, annealing far from the unpeeled front is suppressed both by the loop factor (if we have chosen $c\neq 0$) and by the fact that the molten bubble thus created is shorter than the corresponding number of unpeeled bases. Note that while the physical origin of these two effects is the same, namely the entropy associated with ssDNA, we deal separately with extensive and nonextensive components.
\bibitem{note} Hysteresis induced by the slow dynamics of melting is also pronounced (for large $c$) if we take an alternative form for the nonextensive entropy of molten regions, $\Delta S_{\rm loop}(n) = - \kb \, c \ln (1+n)$~\cite{bubbles1}.
\end{thebibliography}
\end{document}